\documentclass{aa}  

\usepackage{graphicx}
 \usepackage[varg]{txfonts}
\usepackage{natbib}
\usepackage[colorlinks=true,citecolor=blue,linkcolor=blue]{hyperref}

\graphicspath{{./figs/}}
\newcommand{\chandra}{\textsl{Chandra}\xspace}

\begin{document}

\title{The clumpy absorber in the high-mass X-ray binary Vela X-1}

   \author{
V.~Grinberg\inst{1}\and
N. Hell\inst{2}  \and 
I.~El~Mellah\inst{3} \and 
J.~Neilsen\inst{4}  \and  
A.A.C.~Sander\inst{5} \and
M.~Leutenegger\inst{6,7} \and 
F.~F\"urst\inst{8}   \and 
D.P.~Huenemoerder\inst{9}  \and 
P.~Kretschmar\inst{8}  \and 
M.~K\"uhnel\inst{10}  \and 
S.~{Mart{\'{\i}}nez-N{\'u}{\~n}ez}\inst{11} \and
S.~Niu\inst{10,12} \and 
K.~Pottschmidt\inst{6,7} \and 
N.S.~Schulz\inst{9}  \and
J.~Wilms\inst{10} \and
M.A.~Nowak\inst{9} }

\institute{ESA European Space Research and
  Technology Centre (ESTEC), Keplerlaan 1, 2201 AZ Noordwijk, The Netherlands\\
  \texttt{vgrinberg@cosmos.esa.int}
  \and
  Lawrence Livermore National Laboratory, 7000 East Avenue, Livermore, CA 94550, USA
  \and 
  Centre for Mathematical Plasma Astrophysics, Department of
  Mathematics, KU Leuven, Celestijnenlaan 200B, 3001 Leuven, Belgium
\and
Villanova University, Department of Physics, Villanova, PA 19085, USA
  \and
  Institut f\"ur Physik und Astronomie, Universit\"at Potsdam,
  Karl-Liebknecht-Str.~24/25, 14476 Potsdam, Germany
\and 
  CRESST, Department of Physics, and Center for Space Science and 
  Technology, UMBC, Baltimore, MD 21250, USA 
  \and 
  NASA Goddard Space Flight Center, Greenbelt, MD 20771, USA 
  \and 
  ESA European Space Astronomy Centre (ESAC), Science Operations 
  Departement, 28692 Villanueva de la Ca\~nada, Madrid, Spain 
  \and 
  Massachusetts Institute of Technology, Kavli Institute for 
  Astrophysics and Space Research, Cambridge, MA 02139, USA 
  \and 
  Dr. Karl Remeis-Sternwarte and Erlangen Centre for Astroparticle
  Physics (ECAP), Universit\"at Erlangen-N\"urnberg, Sternwartstrasse
  7, 96049 Bamberg, Germany 
  \and 
Instituto de F\'isica de Cantabria (CSIC-Universidad de Cantabria), 39005 Santander, Spain
  \and 
 Shanghai Astronomical Observatory, Chinese Academy of Sciences, 80 Nandan Road, Shanghai, 200030, China  
}

\date{Received 28 August 2017/ Accepted 17 November 2017}

\abstract{Bright and eclipsing, the high-mass X-ray binary
  Vela X-1 offers a unique opportunity to study accretion
  onto a neutron star from clumpy winds of O/B stars and to
  disentangle the complex accretion geometry of these systems. In
  \emph{Chandra}-HETGS spectroscopy at orbital phase $\sim$0.25, when
  our line of sight towards the source does not pass through the
  large-scale accretion structure such as the accretion wake, we
  observe changes in overall spectral shape on timescales of a few
  kiloseconds. This spectral variability is, at least in part, caused
  by changes in overall absorption and we show that such
  strongly variable absorption cannot be caused by unperturbed clumpy
  winds of O/B stars.
  We detect line features from high and low ionization species of
  silicon, magnesium and neon whose strengths and presence
  depend on the overall level of absorption. They imply a co-existence
  of cool and hot gas phases in the system that we interpret as a
  highly variable, structured accretion flow close to the compact
  object such as has been recently seen in simulations of wind
  accretion in high-mass X-ray binaries.}


   \maketitle

\section{Introduction}

The high-mass X-ray binary (HMXB) \object{Vela X-1} consists
of a $\sim$283\,s period pulsar \citep{McClintock_1976a} in an
eclipsing $\sim$9\,d orbit with the B0.5~Ib supergiant HD~77581
\citep{Hiltner_1972a,Forman_1973a}. The radius of \object{HD 77581} is
30\,$R_{\odot}$ and the binary separation 53.4\,$R_{\odot}$ or
$\sim$1.7\,$R_\mathrm{HD~77581}$
\citep{van_Kerkwijk_1995a,Quaintrell_2003a} so that the neutron star
is deeply embedded in the strong \citep[mass-loss rate
${\sim}2\times10^{-6}\,M_\odot\,\mathrm{yr}^{-1}$,][]{Watanabe_2006a}
wind of the companion that it accretes from.
Newest estimates consistently point towards a low terminal velocity,
$v_\infty$, of 600--750\,$\mathrm{km}\,\mathrm{s}^{-1}$
\citep{Krticka_2012a,Gimenez-Garcia_2016a,Sander_2017b}. 
Despite having a modest average X-ray luminosity
of ${\sim}4\times 10^{36}\,\mathrm{erg}\,\mathrm{s}^{-1}$, with a
distance of $\sim$2\,kpc \citep{Nagase_1986a} Vela X-1 is close enough
that it is one of the brightest HMXBs in the sky. 

The high inclination of the system \citep[$> 73^\circ,$][]{Joss_1984a}
naturally allows probing of the accretion and wind geometry
through observations at different orbital phases
\citep[e.g.,][]{Haberl_1990a,Goldstein_2004a,Watanabe_2006a,Fuerst_2010a}.
Outside of the eclipse, the base level of absorption in the system
shows a systematic evolution along the binary orbit, with average
column densities ranging from a few times $10^{22}\,\mathrm{cm}^{-2}$
at $\phi_\mathrm{orb} \approx 0.25$ to
$15 \times 10^{22}\,\mathrm{cm}^{-2}$ at
$\phi_\mathrm{orb} \approx 0.75$ \citep{Doroshenko_2013a}. The
evolution does not agree with expectations for an unperturbed wind and
is attributed to the changing sightline through the complex accretion
geometry of the system that includes an accretion wake, a
photoionization wake, and possibly a tidal stream
\citep[e.g.,][]{Eadie_1975a,Nagase_1983a,Sato_1986a,Blondin_1990a,Kaper_1994a,
  Malacaria_2016a}.

Superimposed on this change are irregular individual absorption events
\citep{Haberl_1990a, Odaka_2013a} and bright flares
\citep{Martinez-Nunez_2014a} of varying length as well as so-called
off-states \citep{Kreykenbohm_2008a,Sidoli_2015a}. Clumpiness of the
companion wind has been invoked by several authors to explain the
short-term variability through absorption in or accretion of clumps
\citep{Kreykenbohm_2008a,Fuerst_2010a,Martinez-Nunez_2014a}, while
\citet{Manousakis_2015a} invoke unstable hydrodynamic
flows. At least the off-states can also be explained
by a quasi-spherical subsonic accretion model
\citep{Shakura_2013a}. The seminal simulations of stellar-wind
disruption by a compact source of \citet{Blondin_1990a} show an
oscillating accretion wake structure that also leads to variability in
the accretion rate.

High-spectral-resolution observations allow us to address the
structure and distribution of material in the system: as plasmas of
different densities, temperatures, and velocities reprocess the
radiation from the neutron star, they imprint characteristic features
onto the X-ray continuum. Previous analyses of the existing
\textsl{Chandra} High Energy Transmission Grating Spectrometer
\citep[\textsl{Chandra-}HETGS;][]{Canizares_2005a} observations
concentrated on the global properties of the plasma: emission lines
from a variety of ionization stages of Ne, Mg, Si, and Fe, including
K$\alpha$ emission from L-shell ions in some cases, have been detected
in eclipse \citep[$\phi_{\mathrm{orb}} \approx 0$,][]{Schulz_2002b}
and when looking through the accretion wake
\citep[$\phi_{\mathrm{orb}} \approx
0.5$,][]{Goldstein_2004a,Watanabe_2006a}. These observations allow us
to determine the physical conditions in this system, including the
properties of the wind. \citet{Schulz_2002b} showed that a hot
optically thin photoionized plasma and a colder plasma that gives rise
to the fluorescent K$\alpha$ lines from L-shell ions,
co-exist in the system. Comparing fluorescent emission lines
at eclipse and at $\phi_{\mathrm{orb}} \approx 0.5$,
\citet{Watanabe_2006a} find them an order of magnitude brighter at
$\phi_{\mathrm{orb}} \approx 0.5$. They thus conclude that the lines
have to be produced in a region that is occulted during the eclipse
and thus lies between neutron star and companion.

Surprisingly, data obtained at $\phi_{\mathrm{orb}} \approx 0.25$,
when our line of sight towards the approaching neutron star is relatively
unobscured (Fig.~\ref{fig:sketch}), proved more difficult to analyze
and interpret. \citet{Goldstein_2004a} describe emission and
absorption lines with a strong continuum contribution. They conclude
that these lines must come from different regions, but do not put
constraints on the location of the material.  \citet{Watanabe_2006a}
do not discuss the spectrum at this orbital phase in detail as they
conclude that it is too continuum dominated for their purposes, but
they mention possible evidence for weak absorption lines. In their 3D
Monte Carlo simulations of X-ray photons propagating through the wind,
the model shows P\,Cygni profiles at
$\phi_{\mathrm{orb}} \approx 0.25$ that they conclude to not be
observable with \textsl{Chandra} given the spectral resolution and
sensitivity.

At the same time, recent results with \emph{Suzaku}
\citep{Odaka_2013a} and \emph{XMM-Newton} \citep{Martinez-Nunez_2014a}
at orbital phases of $\sim$0.25 show strong changes in absorption on
time scales as short as $\sim$ks. Similar behavior has also been
indicated in historical observations with \textsl{Tenma} and
\textsl{EXOSAT} \citep{Nagase_1986a,Haberl_1990a}. However, the
\textsl{Chandra} observations of Vela X-1 have previously been only
analyzed in a time averaged way, disregarding possible spectral
changes within a given observation. This shortcoming prompted us to
re-visit an early $\sim$30\,ks long \chandra-HETGS observation of Vela
X-1 at $\phi_{\mathrm{orb}} \approx 0.25$ with the aim to search for
variability on shorter timescales using \chandra's
high-resolution capabilities.

We first introduce our data, discuss the overall variability during our observation and motivate the division of the total exposure into periods of low and high hardness, which correspond to low and high absorption and are to be analyzed separately, in Section~\ref{sec:data}. In Section~\ref{sec:lines}, we analyze the line content of the resulting spectra. In Section~\ref{sec:discussion}, we discuss the observations with a focus on the wind structure and the interactions of the compact object with the wind. We close with a summary and an outlook towards future observations and laboratory work necessary to further our understanding of the structured wind and clumpy absorber in Vela X-1 in Sec.~\ref{sec:sum}.

\section{Data and variability behavior}\label{sec:data}

\begin{figure}
\resizebox{\hsize}{!}{\includegraphics{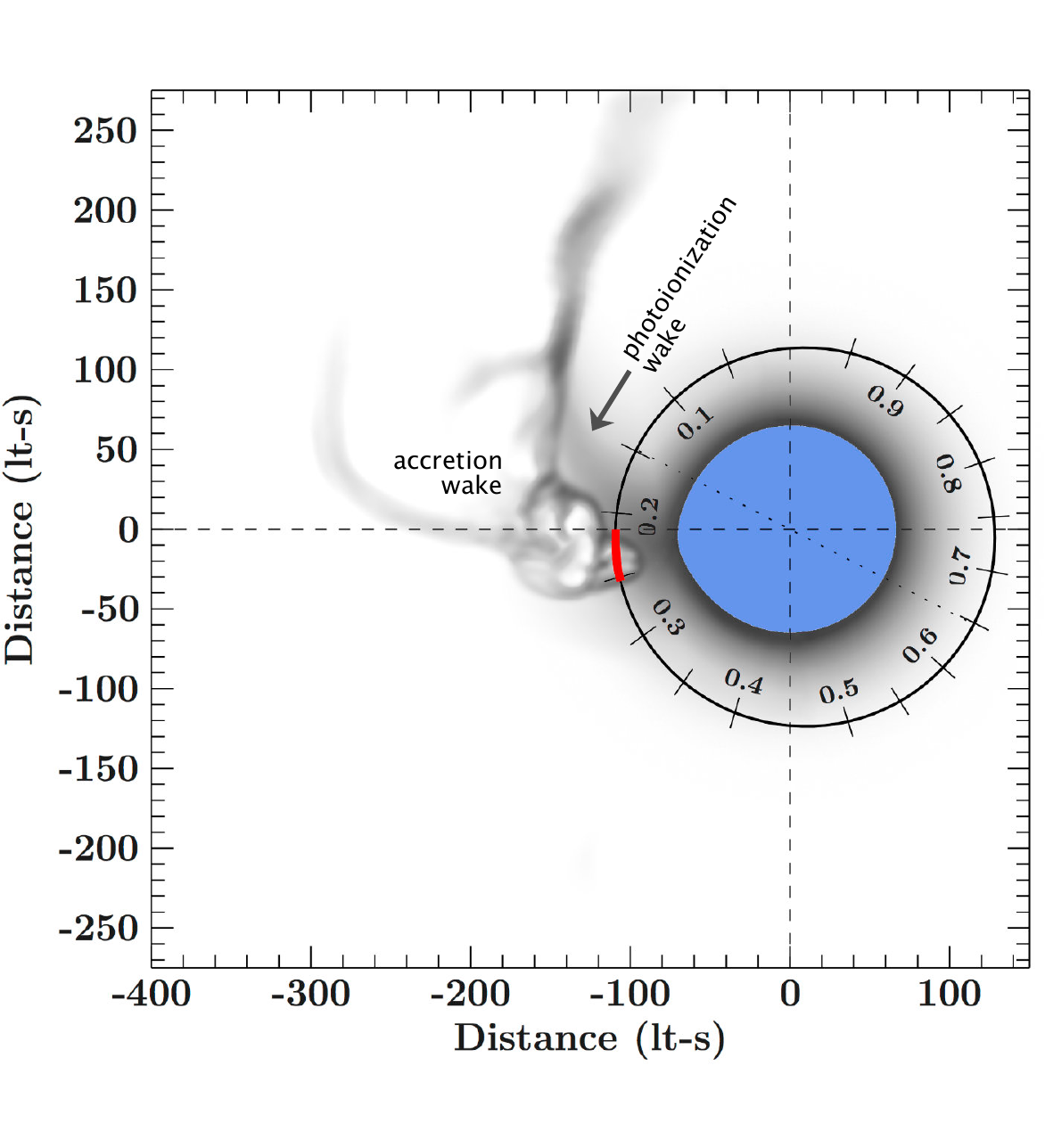}}
\caption{To-scale sketch of the Vela X-1 system. HD~77581 shown in
  blue, the orbit of the neutron star is represented by the black
  ellipse. Orbital phases are marked and labelled; phases covered in
  the present work are shown in red. We emphasize that the orbital
  phases are equidistant in time for the present orbital shape. We
  assume $i = 73^{\circ}$, i.e., lower limit for inclination according
  to \citet{Joss_1984a}, and the distances are thus lower limits. The
  background shows an indicative image of the variable large-scale
  structures that are formed through the interactions of the neutron
  star with the stellar wind of the mass donor. This image is an
  artistic impression based on simulations published in
  \citet{Manousakis_2011_PhD}. The observer is towards minus infinity
  along the y-axis.}\label{fig:sketch}
\end{figure}

We use the ObsID~1928 taken with \chandra-HETGS on 2001-Feb-05 for
$\sim$30\,ks at orbital phases $\phi_{\mathrm{orb}} = 0.21$--0.25
\citep[according to the ephemeris of][where
$\phi_{\mathrm{orb}} = 0$ is defined as mid-eclipse, Fig.~\ref{fig:sketch}]{Kreykenbohm_2008a}. This ObsID has previously been included in
the analyses of \citet{Goldstein_2004a}, who showed that the spectrum
is not affected by pile-up, and \citet{Watanabe_2006a}; both extract
one spectrum for the full ObsID only and \citet{Watanabe_2006a} do not
analyze the line content of this observation.

We follow the standard \textsl{Chandra} threads and use CIAO~4.7 using
CalDB~4.6.7, but employ a narrower sky mask to improve the flux at
very short wavelengths by avoiding some of the overlap
between extraction regions. For bright sources, such as X-ray binaries
in general and Vela X-1 in particular, \textsl{Chandra}-HETG
observations can be considered as essentially background-free. All
analyses in this paper were performed with the Interactive Spectral
Interpretation System (ISIS)~1.6.2
\citep{Houck_Denicola_2000a,Houck_2002,Noble_Nowak_2008a}.
Uncertainties are given at the 90\% confidence level for one parameter
of interest.

\subsection{Light curves and hardness}

\begin{figure}
\resizebox{\hsize}{!}{\includegraphics{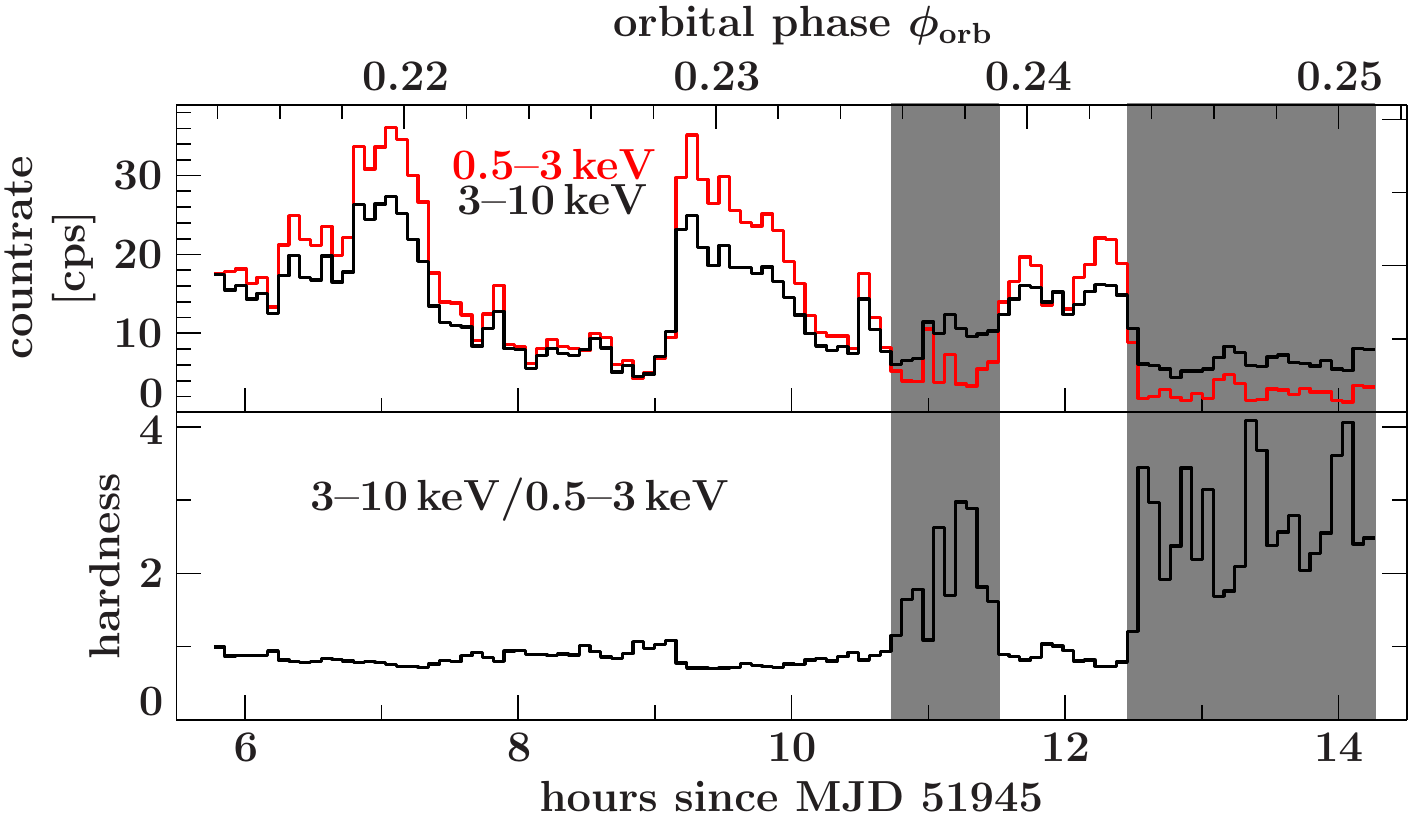}}
\caption{Light curves and hardness ratio of Vela X-1 (\textsl{Chandra}
  ObsID~1928) in the 0.5--3\,keV and 3--10\,keV energy bands, binned
  to $\sim$283\,s. The hardness is defined as the quotient of the
  count rates in the 3--10\,keV and 0.5--3\,keV bands. Gray shaded
  areas indicate the time periods used to extract the high hardness
  spectrum, unshaded regions are used for the low hardness
  spectrum.}\label{fig:lc}
\end{figure}

Figure~\ref{fig:lc} shows light curves extracted with the approximate
time resolution of the neutron star spin period (283\,s) in the
0.5--3\,keV and 3--10\,keV energy bands and the corresponding
hardness, defined as the ratio between the harder and the softer
band. We can distinguish two distinct behaviors: periods of highly
variable high hardness and of stable low hardness. Interestingly, the
total count rate changes by a factor of a few during the period of
stable low hardness. Such increases in flux can be attributed to
increases in the intrinsic X-ray brightness of the neutron star
\citep[e.g.,][]{Martinez-Nunez_2014a}. To investigate the origin of
the changing hardness, we time-filtered our observation and
extracted one spectrum each for the low and high hardness periods
using user-defined good time intervals corresponding to the two
periods highlighted in Fig.~\ref{fig:lc}. Exposures and total counts
are listed in Table~\ref{tab:gtis}.

\begin{table}
\caption{Exposure and total counts in
    hardness-resolved spectra}\label{tab:gtis}     
\centering                                     
\begin{tabular}{l c c}          
\hline\hline                      
hardness & exposure & total counts \\
\hline                                  
  high & 9\,ks & $96.7\times10^3$ \\  
  low & 20\,ks & $630\times10^3$ \\
\hline
\end{tabular}
\end{table}

\subsection{Continuum variability}\label{sec:cont}

We first address changes in the overall continuum shape between low and high hardness periods. A physical model for X-ray continuum emission in accretion-powered X-ray pulsars is an open question \citep[but see][for advanced approaches]{Becker_2007a,Farinelli_2016a,Wolff_2017a}. Thus a multitude of empirical models, usually some combination of absorbed power laws, has been successfully used to describe spectra of accreting pulsars \citep[e.g.,][]{Mueller_2013a}.

\begin{figure}
\resizebox{\hsize}{!}{\includegraphics{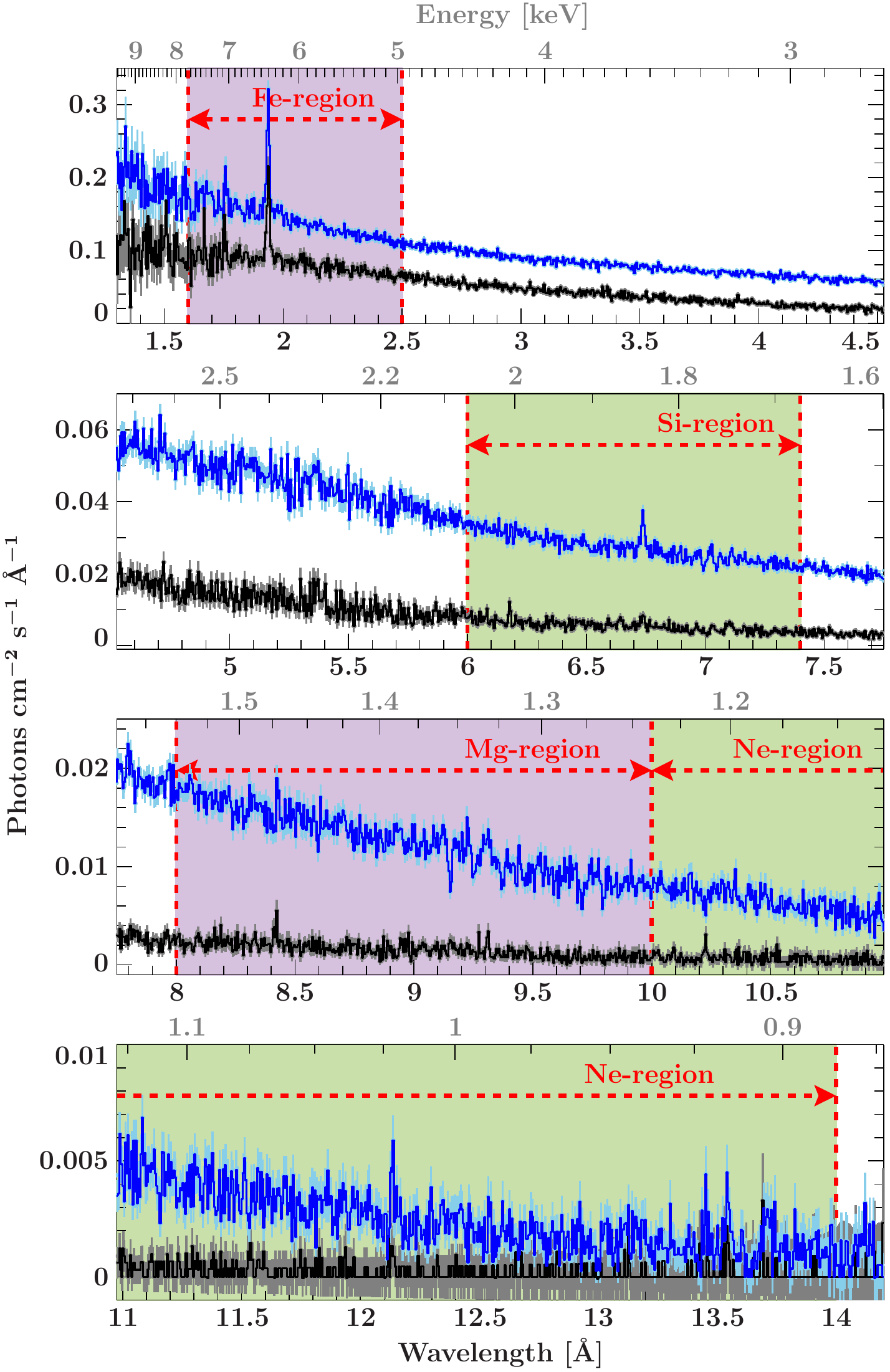}}
\caption{Vela X-1 spectra from ObsID 1928 from low (blue) and high
  (black) hardness phases, HEG and MEG combined and rebinned by a
  factor of 2. Red dashed lines indicate the borders of the four
  regions (shaded) used in the local power law fits in
  Sect.~\ref{sec:lines}}\label{fig:zoom}
\end{figure}

In Vela~X-1, the presence of additional possibly X-ray emitting and
reprocessing components such as the stellar wind, the denser material
in the accretion and ionization wakes as well as the surface of
HD~77851, makes modelling the soft X-ray emission especially
challenging (see Fig.~\ref{fig:zoom} for an overview of the
0.8--9.5\,keV spectrum). Below $\sim$2.5\,keV, the spectrum is complex
and shows strong line features and peculiar continuum components such
as a soft excess
\citep{Nagase_1986a,Hickox_2004a,Martinez-Nunez_2014a} that depend on
orbital phase and absorption \citep{Haberl_1990a}.  We therefore
define the continuum based on the 2.5--10\,keV band, which does not
include any strong lines with the exception of the Fe~K$\alpha$
complex \citep{Torrejon_2010a,Gimenez-Garcia_2015a}, which we exclude
by ignoring the 6.2--6.5\,keV band.  To improve the signal-to-noise
ratio, we combine the positive and negative first-order High
Energy Grating (HEG) and Medium Energy Grating (MEG) data after
rebinning to the MEG grid. We then rebin the combined spectrum to a
SNR of 5 and a minimum of four MEG channels per bin.

We describe the high and low hardness spectra each with a power law,
defined by a power law index $\Gamma$ and a normalization, that is
modified by absorption modelled with \texttt{tbnew}, the new version of \texttt{tbabs}
\citep{Wilms_2000a}, using \texttt{wilm} abundances
\citep{Wilms_2000a} and \texttt{vern} cross-sections
\citep{Verner_1996a}.
The assumed model is simple and will not result in a
perfect description of the continuum, but given the quality of the
data it is adequate for our aim to assess continuum-shape changes.

\begin{figure}
\resizebox{\hsize}{!}{\includegraphics{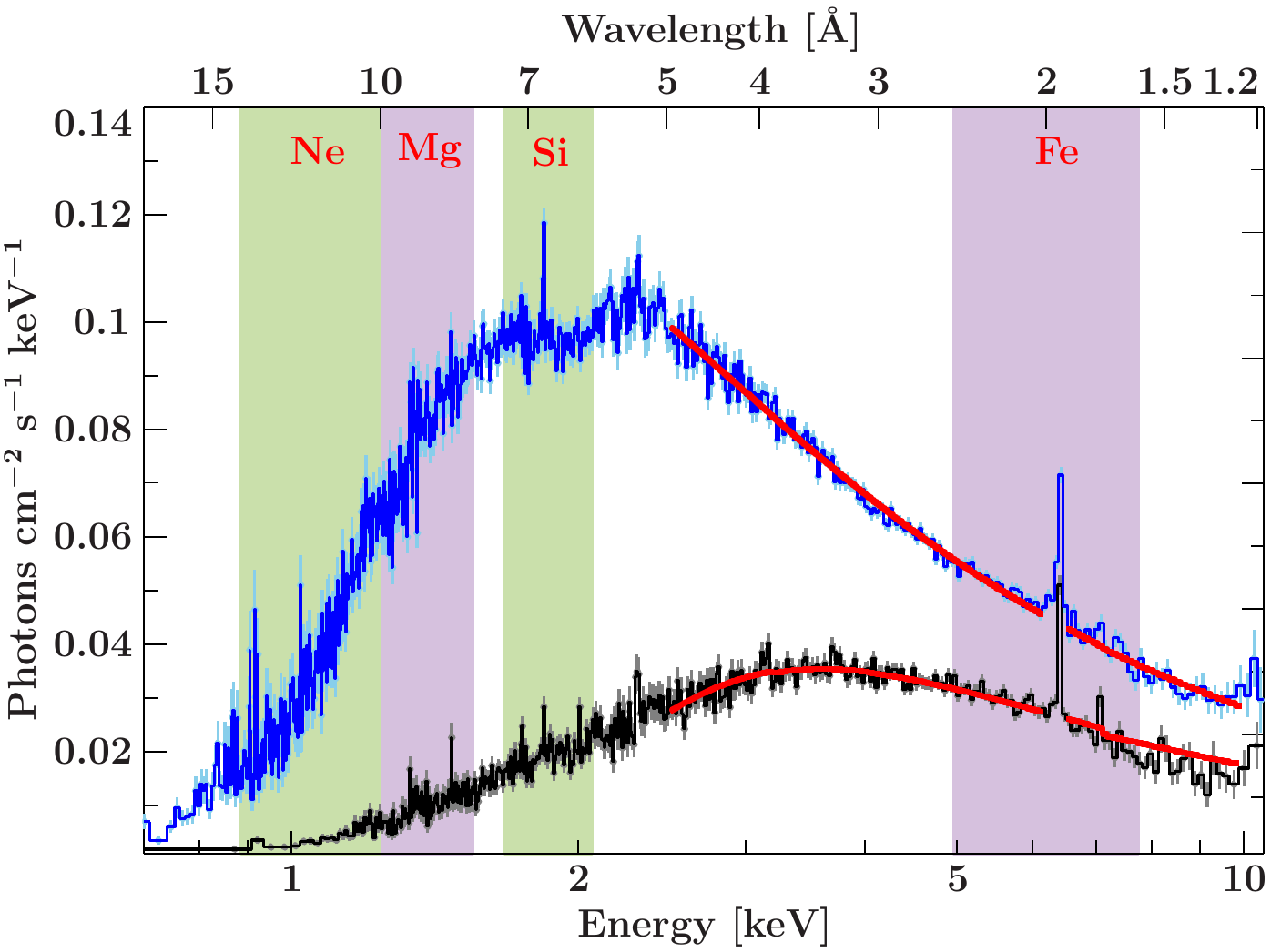}}
\caption{\chandra-HETGSs spectra of Vela X-1 from ObsID 1928
  from low (blue) and high (black) hardness periods. The best fitting
  absorbed power-law model with the same photon index, but varying
  normalization and absorption between the high- and
  low-hardness
  spectra is shown in red in the modelled range. Shaded areas indicate
  the four regions used in the local power-law fits in
  Sect.~\ref{sec:lines}}\label{fig:continuum}
\end{figure}

To assess whether changes in absorption, expressed through the
equivalent hydrogen column density, $N_{\mathrm{H}}$, or in
power-law slope, $\Gamma$, lead to the hardness changes, we
perform two fits, with results listed in Table~\ref{tab:continuum}.
In a first fit, we assume that the absorption column, described by the
$N_{\mathrm{H}}$ parameter, is the same for both low and high hardness
and tie this parameter between the two hardness modes.  No acceptable
fit can be obtained in this case.  In a second approach, we
assume the same power-law shape, i.e., we tie $\Gamma$, but
let the absorption be different. This model results in a better
description of the data (Fig.~\ref{fig:continuum}).

\begin{table*}
\centering \caption{Results of absorbed power law fits to the
  2.5--10\,keV continuum. The subscript ``h'' and ``l'' refer to model
  components applied to spectra taken during high- and
  low-hardness periods, respectively. We list best-fit
  results for three cases: no ties between parameters describing
  low- and high-hardness spectra (``no ties''), same
  equivalent hydrogen column width of absorption for low- and
  high-hardness spectra (``tied $N_{\mathrm{H}}$''), and same photon index
  for low- and high-hardness spectra (``tied
  $\Gamma$'').}\label{tab:continuum} \renewcommand{\arraystretch}{1.3}
\begin{tabular}{lccccccc}
\hline
\hline
 & $N_{\mathrm{H,h}}$ & norm$_{\mathrm{h}}$ (at
1\,keV) & $\Gamma_\mathrm{h}$ & $N_{\mathrm{H,l}}$ &
                                                                       norm$_{\mathrm{l}}$ (at
1\,keV)
  & $\Gamma_{\mathrm{l}}$ & $\chi^2$/dof \\
& [$10^{22}\,\mathrm{cm}^{-2} $] &
                                                           [keV$^{-1}$\,cm$^{-2}$\,s$^{-1}$]
  & & [$10^{22}\,\mathrm{cm}^{-2} $] &
                                                           [keV$^{-1}$\,cm$^{-2}$\,s$^{-1}$]
\\
\hline
tied $N_{\mathrm{H}}$   & $1.68\pm0.18$ &  $0.072\pm0.004$ & $0.507\pm0.030$ & =
                                                               $N_{\mathrm{H,h}}$
                                         &
                                                              $0.349\pm0.017$  & $1.115\pm0.027$
               & $955/354$\\
tied $\Gamma$ & $5.76\pm0.25$ & $0.178\pm0.009$ & $0.989\pm0.026$ &
                                                              $0.72\pm0.18$
                                                              & $0.277\pm0.013$&
                  =$\Gamma_{\mathrm{h}}$ & $473/354$
 \\
 no ties & $6.7\pm0.5$ &  $0.23\pm0.03$ & $1.13\pm0.06$ &
                                                              $0.52\pm0.19$
                                                              & $0.261\pm0.014$&
                 $0.956\pm0.029$ & $454/353$
 \\
\hline
\end{tabular}
\end{table*}

\begin{figure}
\resizebox{\hsize}{!}{\includegraphics{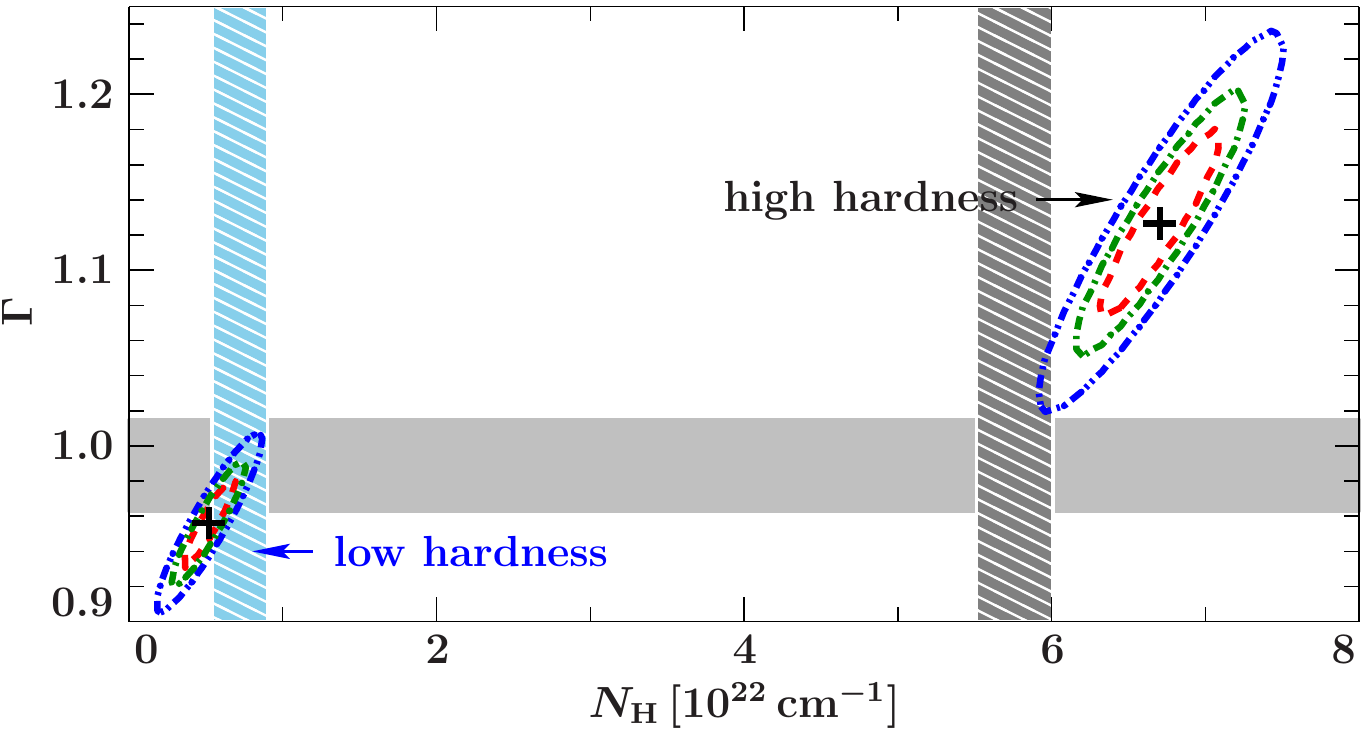}}
\caption{Confidence intervals for independent power-law fits
  to low and high hardness data; crosses represent the best fit. The
  contours correspond to 1$\sigma$ (red, dashed), 90\% (green,
  dash-dotted) and 99\% (blue, dash-dot-dotted) confidence
  intervals. The solid grey shaded region represents the
  values of $\Gamma$ allowed when the power law slope is tied between
  the low- and high-hardness spectra; the light blue and grey
  patterned areas represent the corresponding allowed ranges for
  $N_{\mathrm H}$ for low and high hardspess
  spectra.}\label{fig:conf_lowhigh}
\end{figure}

Since there is still some systematic deviation between data and model
above $\sim$7\,keV for the high hardness spectrum, we also fit the two
spectra with all model parameters independent of each other
(Table~\ref{tab:continuum}). In Fig.~\ref{fig:conf_lowhigh}, we show
the confidence intervals for $\Gamma$ and $N_{\mathrm H}$ for this
case as well as the 90\% allowed ranges for $\Gamma$ and
$N_{\mathrm H}$ for the case when the power law slope is tied between
the two spectra. The confidence contours show the well known
degeneracy between the derived values of $\Gamma$ and $N_{\mathrm H}$
\citep[e.g.,][]{Suchy_2008a}. We also see that letting $\Gamma$ free
leads to an even larger difference in hydrogen column density between
the two spectra than when $\Gamma$ is tied.  The bulk of the changes
in overall spectral shape are thus due to absorption and not due to
changes in the shape of the underlying power law continuum, although
some changes in spectral shape may also be present.

\section{High-resolution spectroscopy}\label{sec:lines}

Having established that the changes in overall continuum shape between the low- and high-hardness periods are driven by changes in absorption, we now turn to high-resolution spectroscopy where we focus on regions below 2.5\,keV (above 5\,{\AA}) as well as on the Fe~K$\alpha$ complex, which were not included in the broadband spectral modelling due to their complexity. We first visually inspect the differences between the low- and high-hardness spectra as presented in Fig.~\ref{fig:zoom} and compare to the results of \citet{Goldstein_2004a}, who analyzed the spectrum of ObsID~1928 without subdividing it into hardness-dependent periods.  Because of the larger exposure and flux of the low hardness data (Table~\ref{tab:gtis}), we expect that this period has dominated their analysis. We note a difference in line contributions between the spectra at different hardnesses: features change their relative strengths (e.g., lines of the He-like \ion{Si}{xiii} triplet at $\sim$6.7\,{\AA}, see Sec.~\ref{sec:si}), change from emission to absorption (e.g, \ion{Ne}{x}\,Ly\,$\beta$ at $\sim$10.25\,{\AA}, see Sec.~\ref{sec:Ne}), and new features appear that were not found in the earlier analysis, such as lines from low ionization stages of Si around $\sim$7\,{\AA} (see Sect.~\ref{sec:si}).

In the remainder of this section, we model the low- and
  high-hardness spectra independently.  As in Sec.~\ref{sec:cont}, we
combine the positive and negative first-order HEG and MEG
data after rebinning to the MEG grid, but do not rebin the spectra
further. Except where explicitly stated otherwise, we do not assume
different line features to originate at the same location or the same
medium, especially since multiple gas phases with different
temperatures have been suggested as explanations for observations at
higher orbital phases \citep{Goldstein_2004a, Watanabe_2006a}. Our
analysis of the lines is based on local power-law continuum fits,
i.e., power-law continua describing a narrow interval around
the respective features of interest. Although the slope, $\Gamma$, and
normalization of such power laws do not have a direct physical
meaning, we list the best fit results obtained for all regions
discusssed in the remainder of this section in the spirit of
reproducibility of results in Table~\ref{tab:pl}. Our choice of
wavelength intervals used for such local fits is informed by the
previous knowledge of lines observed in Vela X-1 at other orbital
phases \citep{Schulz_2002b,Goldstein_2004a,Watanabe_2006a}.  We
consistently use Cash statistics \citep{Cash_1979a} in this section as
several of the considered intervals require it. We always first fit
for the width of any component.  However, since none of the detected
Gaussian emission components are resolved at MEG resolution
(0.023\,{\AA} FWHM), we fix their widths, $\sigma$, to
$\sigma = 0.003$\,{\AA}, i.e., about one third of the MEG resolution.

\begin{table*}
\centering
\caption{Parameters for the local power-law models used in Sec.~\ref{sec:lines}}\label{tab:pl}
\renewcommand{\arraystretch}{1.3}
\begin{tabular}{llcccc}
\hline
\hline
wavelength & region & \multicolumn2c{high hardness} & \multicolumn2c{low hardness} \\
range & & $\Gamma$ & norm  (at 1\,keV) & $\Gamma$ &
                                                                norm  (at 1\,keV) \\
$[${\AA}$]$& && [keV$^{-1}$\,cm$^{-2}$\,s$^{-1}$] & &
                                        [keV$^{-1}$\,cm$^{-2}$\,s$^{-1}$] \\
\hline
1.6--2.5&Fe~K$\alpha$ & $0.889^{+0.007}_{-0.011}$&$0.137^{+0.028}_{-0.033}$ &
                                                                 $0.846^{+0.004}_{-0.006}$
           & $0.217^{+0.030}_{-0.007}$ \\

6.0--7.4& Si & $-1.17^{+0.31}_{-0.26}$&
                                    $\left(9.13^{+0.23}_{-0.72}\right)\times10^{-3}$& $0.067\pm0.008$ & $0.1009^{+0.0012}_{-0.0010}$\\
8.0--10& Mg (individual lines)& $-3.21^{+0.09}_{-0.44}$ &
                                     $\left(3.2\pm0.5\right)\times10^{-3}$
& $-1.575\pm0.020$ & $0.0477^{+0.0005}_{-0.0003}$ \\
8.0--10&Mg (template) & -- & -- & $-1.562\pm0.020$ & $0.0479^{+0.0011}_{-0.0014}$\\
10--14&Ne & $-4.6^{+0.8}_{-0.9}$& $\left(2.51^{+0.30}_{-0.28}\right)\times10^{-3}$ & $-4.41\pm0.09$& $0.0271^{+0.0008}_{-0.0006}$\\
\hline
\end{tabular}
\end{table*}

We use two approaches to search for lines: a blind search based on the
Bayesian Blocks algorithm\footnote{As implemented in the
  \texttt{SITAR} package
  (\url{http://space.mit.edu/cxc/analysis/SITAR/}) and included in the
  \texttt{isisscripts}
  (\url{http://www.sternwarte.uni-erlangen.de/isis/}).}
\citep{Scargle_2013a} and a conventional, manual approach, where we
utilize our knowledge of expected line energies of a given
element. The Bayesian Blocks approach for spectral data is
introduced, discussed, and benchmarked against other methods in
\citet{Young_2007a}. With this latter approach, we test for
the existence of lines against a continuum model, starting with a
power law that we iteratively fit to the data. At each iteration we
extend the model by adding any lines found in the previous step.

The algorithm treats the data in an unbinned manner, but compares the
changes in likelihood compared to the applied model\footnote{In the
  original algorithm as applied to timing data, the model is an
  assumed constant rate.  Here the model is a smooth spectral
  continuum.} for grouping the data into different numbers of ``blocks'', with a single block corresponding to the absence of lines.  For a given penalty factor the algorithm exactly solves the $2^{n-1}$ problem, where $n$ is the number of data bins, of how many blocks to divide the data into.  A new block change point can start (or not) between any two bins, hence the $2^{n-1}$ possible divisions of the spectrum. A line is considered to exist between two such change points. Implementations of the Bayesian Blocks algorithm introduce a penalty factor that imposes a requirement for a minimum increase of the log likelihood to consider significance of the addition of a change point to the description of the function (e.g., spectrum or lightcurve). This penalty factor is controlled by an input parameter denoted by $\alpha$ in the work of \citet{Young_2007a} and by $\gamma$ in notation of \citet{Scargle_2013a}. The increase of the log likelihood is approximately equivalent to requiring that the addition of a change point exceeds a significance of $p \sim \exp(-\alpha)$. Since a line requires two such change points, choosing an input of $\alpha=1.5$ is requiring that lines have $p < \exp(-2 \alpha) \approx 0.05$, i.e., they are at least 95\% significant, i.e., $P > 1 - \exp(-2 \alpha) \approx 0.95$.

Our analysis closely follows that of \citet{Young_2007a} where we
start with a large value of $\alpha$, apply the algorithm, and
decrease $\alpha$ in steps of 0.1 until a line is found. We stop at a
lower limit of $\alpha=1.5$.  We list in Tables~\ref{tab:Sihard},
\ref{tab:Mgem}, \ref{tab:Mglow}, \ref{tab:Ne} and \ref{tab:Fe} the
value of $\alpha$ at which (and below) the indicated line is found and
call it $\alpha_{\mathrm{sig}}$.  The line significance is roughly
interpretable as $P\approx1-\exp(-2\alpha_{\mathrm{sig}})$,
  but note that for example He-like triplets are often detected as
one excess encompassing all three lines.

\subsection{Silicon region}\label{sec:si}
We first address the silicon line region (6.0--7.4\,{\AA}). 
We point out that new laboratory reference data are available for the
K-shell transitions of L-shell ions of Si, enabling robust velocity
measurements \citep{Hell_2013a,Hell_2016a}.

\subsubsection{High-hardness spectrum of the silicon region}\label{sec:sihigh}

The Bayesian block approach finds three emission features, all with
$\alpha_{\mathrm{sig}} > 3.3$. We identify these lines with H-like \ion{Si}{xiv}
Ly$\alpha$\footnote{Members of the Ly series in H-like ions commonly
consist of two strong transitions, namely 
$n\mathrm{p}_{3/2}\,{}^2\mathrm{P}_{3/2}$--$1\mathrm{s}_{1/2}\,{}^2\mathrm{S}_{1/2}$
and
$n\mathrm{p}_{1/2}\,{}^2\mathrm{P}_{1/2}$--$1\mathrm{s}_{1/2}\,{}^2\mathrm{S}_{1/2}$,
where the principle quantum number $n=2,3,4,{\ldots}$ of the upper
level corresponds to Ly$\alpha$, Ly$\beta$, Ly$\gamma$, {\ldots}
respectively. Since these two transitions are unresolved in our
spectra, in all cases we use their average wavelength, weighted by a
factor 2:1 according to their statistical weights, as the reference
wavelength.}, He-like \ion{Si}{xiii}~f forbidden line
$(1\mathrm{s}2\mathrm{s}\,{}^3\mathrm{S}_1$--$1\mathrm{s}^2\,{}^1\mathrm{S}_0$,
also known as line z in the notation of \citealt{Gabriel_1972a})
and a blend of the K$\alpha$ lines from \ion{Si}{ii}--\ion{Si}{iv}. We
model these features with Gaussians.

\begin{figure}
\resizebox{\hsize}{!}{\includegraphics{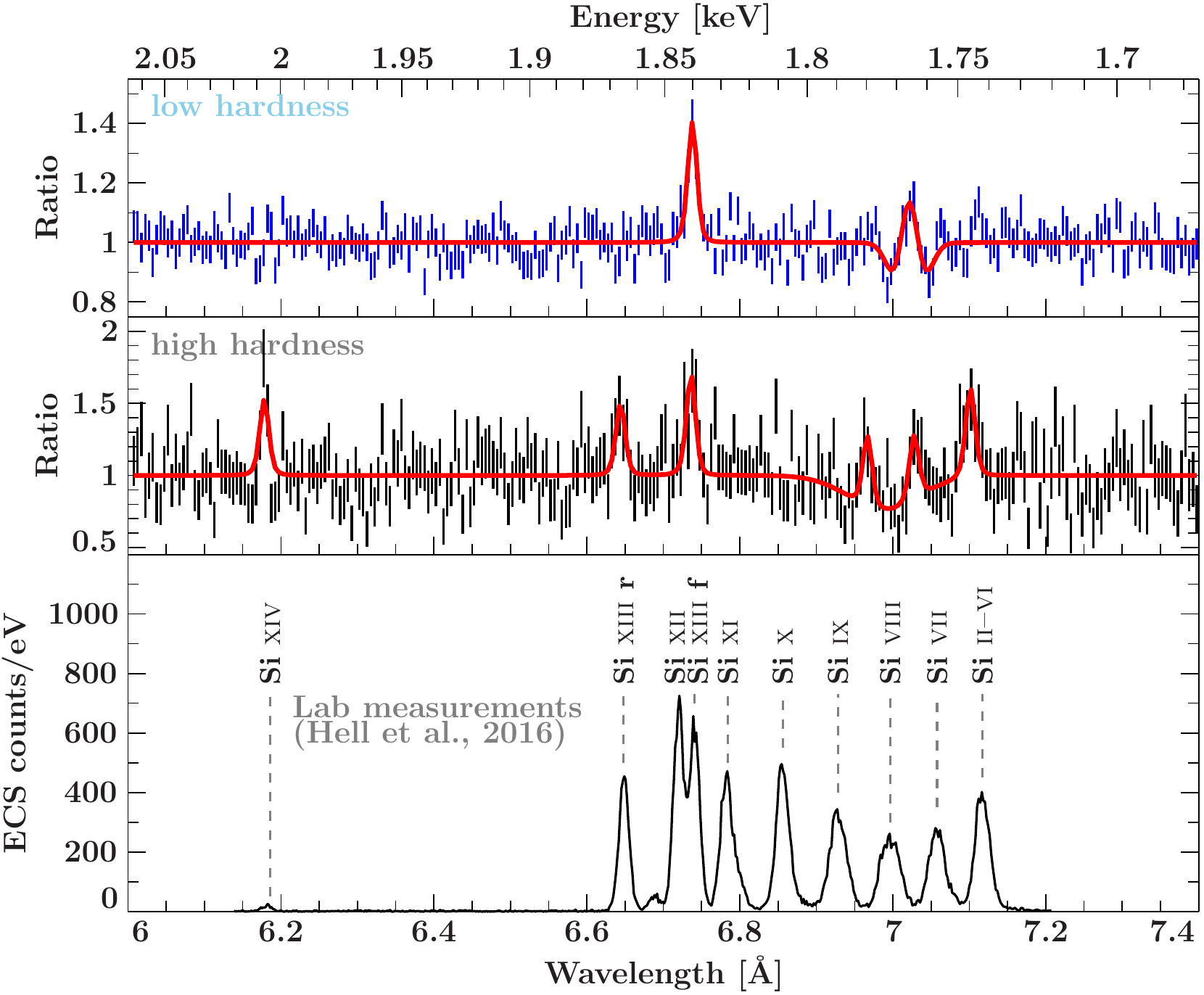}}\hfill
\caption{Silicon region in the low- (upper panel, blue) and
high-hardness (middle panel, black) spectra; both panels show combined
  HEG and MEG $\pm$ first order spectra normalized to their respective
  power-law continua. The solid red line shows the best model fits to
  the data. Residuals are shown in Fig.~\ref{fig:Sispec} The lower
  panel shows the calibrated and summed NASA/GSFC EBIT Calorimeter
  Spectrometer (ECS) silicon spectrum used for line
  identification. These data, as analyzed in \citet{Hell_2016a}, are
  used as best currently available laboratory reference values for the
  energies of K-shell transitions in L-shell ions of
  Si.}\label{fig:Siratios}
\end{figure}

A visual inspection of the residuals, however, reveals further
wave-shaped structures blueward of both
\ion{Si}{xiii}~f and the \ion{Si}{ii}--\ion{Si}{iv} blend. We model these
structures with one and two Gaussian lines, respectively, and identify
them with the He-like \ion{Si}{xiii}~r resonance line
($1\mathrm{s}2\mathrm{p}\,{}^1\mathrm{P}_1$--$1\mathrm{s}^2\,{}^1\mathrm{S}_0$,
also known as line w in the notation of \citealt{Gabriel_1972a}) and
blends of K$\alpha$ transitions in N-like \ion{Si}{viii} and O-like
\ion{Si}{vii} (Fig.~\ref{fig:Siratios}).

In our best fit model that uses six Gaussian emission lines,
large-scale residuals that encompass the area around $\sim${7\,{\AA}}
are still present. We model them with a broad Gaussian ($\sigma
\approx 0.05$\,{\AA}) in absorption that we explicitly do not
interpret as a true line component but as complex continuum shape that
a single power law fails to describe. The introduction of the broad
Gaussian does not significantly change the obtained line positions. We
note especially that the Bayesian Blocks approach relies on a good
description of the continuum; the complex continuum shape therefore
contributed to the failure of the automated approach to find the
\ion{Si}{xiii}~r line and the K$\alpha$ lines of \ion{Si}{vii} and
\ion{Si}{viii}. We do not re-run the Bayesian Blocks algorithm on a
continuum including the additional curvature that we
modelled, as this would be inconsistent with a blind search.

\begin{table*}
\centering 
\caption{Detected silicon emission lines}\label{tab:Sihard}
\renewcommand{\arraystretch}{1.3}
\begin{tabular}{lllcccc}
\hline \hline 
& ion&line & reference wavelength & Bayesian & $v$ & line flux \\
& & & [\AA]& blocks $\alpha_{\mathrm{sig}}$ & $[\mathrm{km}\,\mathrm{s}^{-1}]$&
$[\mathrm{ph}\,\mathrm{s}^{-1}\,\mathrm{cm}^{-2} \times 10^{-5}]$\\ 
\hline 
high & H-like & \ion{Si}{xiv} Ly$\alpha$ & $6.18224\pm
                                    (3.0\times10^{-5})$\tablefootmark{a}
                           & $\geq3.3$ & $-170\pm80$& $5.9^{+2.3}_{-2.2}$\\
hardness &  He-like & \ion{Si}{xiii}~r &
                               6.64795\tablefootmark{b}  & no \tablefootmark{d} & $= v_{\mathrm{Si~\textsc{xiv}}}$&
                                                                                                  $4.5^{+2.2}_{-2.0}$\\
& He-like & \ion{Si}{xiii}~f & 6.74029\tablefootmark{b} &$\geq3.3$ & $=
                                                             v_{\mathrm{Si~\textsc{xiv}}}$&
                                                                                            $6.3^{+2.3}_{-2.2}$\\
& N-like & \ion{Si}{viii} K$\alpha$ & 7.00076\tablefootmark{c} & no \tablefootmark{d} &
                                                                    $-1450\pm200$& $3.8\pm1.7$\\
& O-like & \ion{Si}{vii} K$\alpha$ & 7.05787\tablefootmark{c} & no \tablefootmark{d} & $=
                                                                   v_{\mathrm{Si~\textsc{viii}}}$&
                                                                                                   $3.5\pm1.8$\\
& Al- to Na-like & \ion{Si}{ii}--\textsc{vi} K$\alpha$&
                                                      7.11722\tablefootmark{c}
                                  & $\geq3.3$ & $-680 \pm 250$&
                                                                $4.6^{+1.9}_{-1.8}$\\
\cline{2-7}
low & He-like& \ion{Si}{xiii}~f & 6.74029\tablefootmark{b} &$\geq8.5$ & $-90\pm60$&
                                                                                            $19\pm3$\\
hardness \\
\hline 
\end{tabular}
\tablefoot{
  \tablefoottext{a}{\citet{Erickson_1977a}.
}
  \tablefoottext{b}{\citet{Drake_1988a}; values obtained from the WebGuide version of 
    \texttt{AtomDB} 3.0.2. \citep{Foster_2012a} using the lab/observed 
    wavelengths listed. \texttt{AtomDB} lists uncertainties of 
    $0.002$\,{\AA} for the \citeauthor{Drake_1988a} values, which are 
    overestimated: \citet{Drake_1988a} do not give estimates for the 
    uncertainty of their calculations, but the measurements of 
    \citet{Beiersdorfer_1989a} suggest an agreement between experiment 
    and theory to better than $0.2$\,m\AA, i.e., a factor 10 smaller than 
    \texttt{AtomDB}.}
  \tablefoottext{c}{$1772.01 \pm 0.09$\,eV (\ion{Si}{viii}), $1756.68 
    \pm 0.08$\,eV (\ion{Si}{vii}) and 
    $1742.03 \pm 0.06$\,eV (\ion{Si}{vi}), with additional 0.13\,eV systematic uncertainty \citet{Hell_2016a}}. 
\tablefoottext{d}{We run line searches for $\alpha \geq 1.5_{\mathrm{sig}}$.}
}
\end{table*}

Fitting just one Doppler shift for all detected Si lines does not
result in a satisfactory model for the narrow features. In the
interest of reducing free parameters, it is reasonable to assume that
the lines with a similar ionization stage originate in the same
medium. The lack of intermediate-ionization lines naturally leads to
two line groups: (1) \ion{Si}{xiv}~Ly$\alpha$, \ion{Si}{xiii}~f, and
\ion{Si}{xiii}~r; and \ion{Si}{viii} and \ion{Si}{vii} K$\alpha$. We
therefore fit one line shift to each of these groups (see
Table~\ref{tab:Sihard} for reference values). Using this approach, we
can obtain a good description of the data.

Although we formally fit a Doppler shift for the Si~\textsc{ii--vi} K$\alpha$ line blend, we do not take it into consideration when tying lines or discussing the overall behavior: the contribution of lines to a blend (and thus the measured center of the blend) depends on the properties of the plasma that are certainly different in laboratory and astrophysical environments \citep{Hell_2016a}. In fact, \citeauthor{Hell_2016a} in their NASA/GSFC electron beam ion trap (EBIT) Calorimeter Spectrometer (ECS) data can distinguish two components in the \ion{Si}{ii}--\textsc{iv} K$\alpha$ lines, whose center they see at $1742.03\pm 0.06$\,eV: a blend of F-like \ion{Si}{vi} K$\alpha$ and Ne-like \ion{Si}{v} K$\alpha$ at $1742.88^{+0.15}_{-0.17}$\,eV and a blend of Al- to Na-like \ion{Si}{ii}--\textsc{iv} K$\alpha$ at $1740.04^{+0.27}_{-0.36}$\,eV. Shifting the center of the full blend to 1740.04\,eV, would result in a change in measured velocity of $>250\,\mathrm{km}\,\mathrm{s}^{-1}$.  A further complication is the possible presence of the \ion{Mg}{xii}~Ly$\beta$ at $7.106155$\,{\AA} or 1744.74\,eV \citep{Erickson_1977a}.

\begin{figure*}
\includegraphics[width=0.49\textwidth]{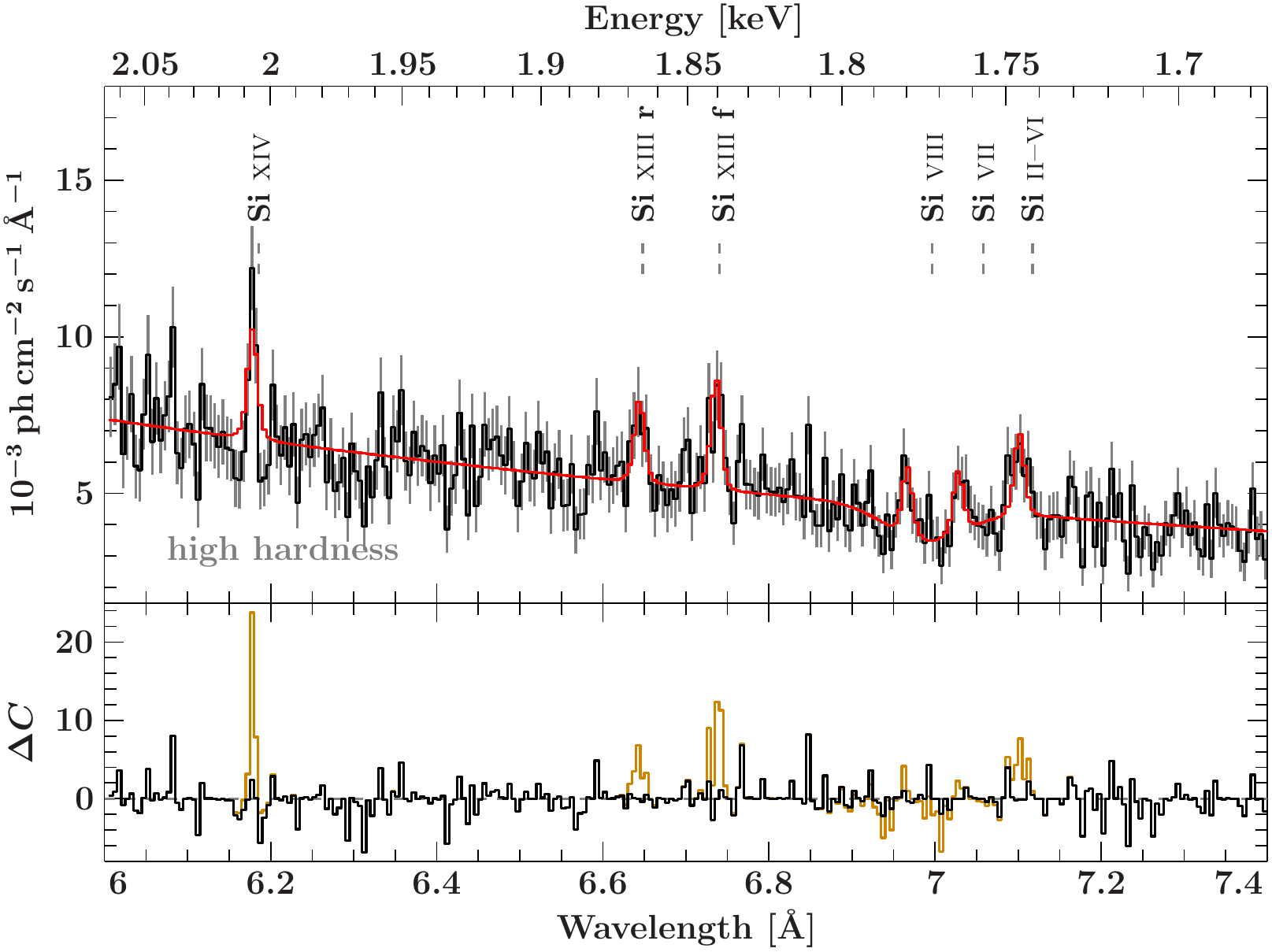}\hfill
\includegraphics[width=0.49\textwidth]{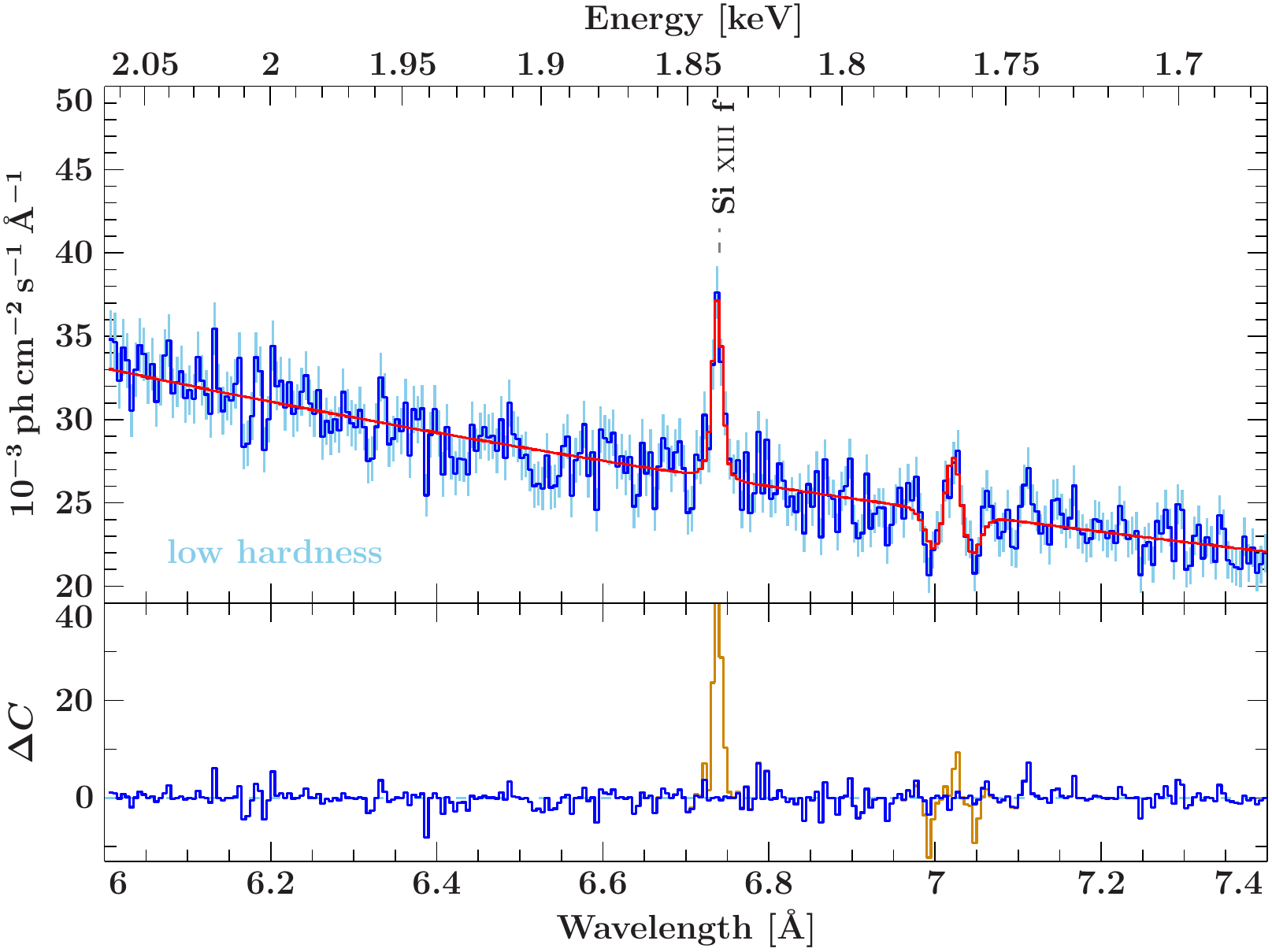}
\caption{Si-region spectra of high hardness (left panel, black) and
  low hardness (right panel, blue) periods. Line identified in
    the spectra are indicated and labelled. Residuals to
  best-fit model are shown in black/blue, residuals to a
  power law without the narrow line components in orange. Spectra
  normalized to their respective power-law continua and
  reference lab measurements are shown in Fig.~\ref{fig:Siratios}.}
\label{fig:Sispec}
\end{figure*}

The resulting properties of the Si lines are summarized in
Table~\ref{tab:Sihard}. The fits are shown in Fig.~\ref{fig:Siratios},
where we show them in comparison with low hardness data and lab
measurements, and in Fig.~\ref{fig:Sispec}, where we also show fit
residuals.

\subsubsection{Low-hardness spectrum of the silicon region}\label{sec:silow}

In the low-hardness spectrum, we find one line with high
significance ($\alpha_{\mathrm{sig}} > 8.5$) at 6.73--6.75\,{\AA} and
a complex w-shaped structure extending 6.98--7.05\,{\AA} at
$\alpha_{\mathrm{sig}} = 1.5$, i.e., our lowest acceptable value of
$\alpha_{\mathrm{sig}}$, with the Bayesian blocks approach.

We identify the Gaussian emission feature as \ion{Si}{xiii}~f line (Table~\ref{tab:Sihard}). The w-structure can be empricially described as a combination of emission and absorption lines at the same energy. This results in lines with widths $\sigma_{\mathrm{em}} \approx \sigma_{\mathrm{abs}} \approx 0.013 \pm 0.002$\,{\AA} at a wavelength of $7.022\pm0.003$\,{\AA} and precludes a line identification (Figs.~\ref{fig:Siratios} and \ref{fig:Sispec}): if this were the Si~\textsc{vii} K$\alpha$ line blend, it would have a Doppler velocity of $\sim$$-$1500\,km\,s$^{-1}$, comparable to the one measured in the high-hardness spectrum. However, no other silicon lines except \ion{Si}{xiii}~f are present in the low hardness spectrum and it would require extreme fine-tuning to produce the \ion{Si}{vii} K$\alpha$ line alone. The peculiar shape of the feature also remains a problem. Regarding other elements, while the Rydberg series of He-like \ion{Mg}{xi}, H-like \ion{Mg}{xii} and H-like \ion{Al}{xiii} have some K$\alpha$ transitions surrounding this region, none of them is sufficiently close or strong enough to explain this feature. Both H-like \ion{Mg}{xii} Ly$\beta$ at 7.10616\,{\AA} and H-like \ion{Al}{xiii}~Ly$\alpha$ at 7.17272\,{\AA} \citep{Erickson_1977a} are too high in wavelength. While the He-like \ion{Mg}{xi} series has several $n \geq 7 \rightarrow 1$ transitions between 7.174\,{\AA} \citep[$n=7$;][]{Verner_1996a} and 7.037\,{\AA} \citep[series limit or ionization potential;][]{Drake_1988a} there is no indication from lines earlier in the series that the lines should be strong enough to explain the observed feature. Still, a radiative recombination continuum (RRC) feature of \ion{Mg}{xi} at 7.037\,{\AA} \citep{Drake_1988a} could contribute to the observed feature.

\subsection{Magnesium region}

We next address the 8.0--10\,{\AA} region, i.e., from longwards in wavelength of \ion{Mg}{xii}~Ly$\alpha$ to shortwards of the \ion{Mg}{iv} K$\alpha$ blend. Several $4 \rightarrow 2$ transitions from \ion{Fe}{xxi} to \ion{Fe}{xxiv} fall into this region, but neither does their location agree with any of the features detected in the following nor do we detect any of the corresponding stronger $3 \rightarrow 2$ transitions from \ion{Fe}{xxi} to \ion{Fe}{xxiv} \citep{Brown_2002a} in Sec.~\ref{sec:Ne}.

\subsubsection{High-hardness spectrum of the magnesium region}\label{sec:Mghigh}

We identify the first line found ($\alpha_{\mathrm{sig}} = 5.4$) by the Bayesian
Block approach with \ion{Mg}{xii} Ly$\alpha$
\citep[$8.42101\pm4.0\times10^{-5}$\,{\AA},][]{Erickson_1977a}. 
The algorithm further finds a very shallow step in the residuals at
$\sim$9.5\,{\AA}, consistent with the location of the magnesium K
edge. However, the step implies an excess where from an edge we would
expect additional absorption 
and consistently, the inclusion of an edge does not improve our model.
Indeed, the best-fit optical depth of the edge component
is zero, i.e., no edge is present, so that we do not use
this component further.

\begin{figure}
\resizebox{\hsize}{!}{\includegraphics{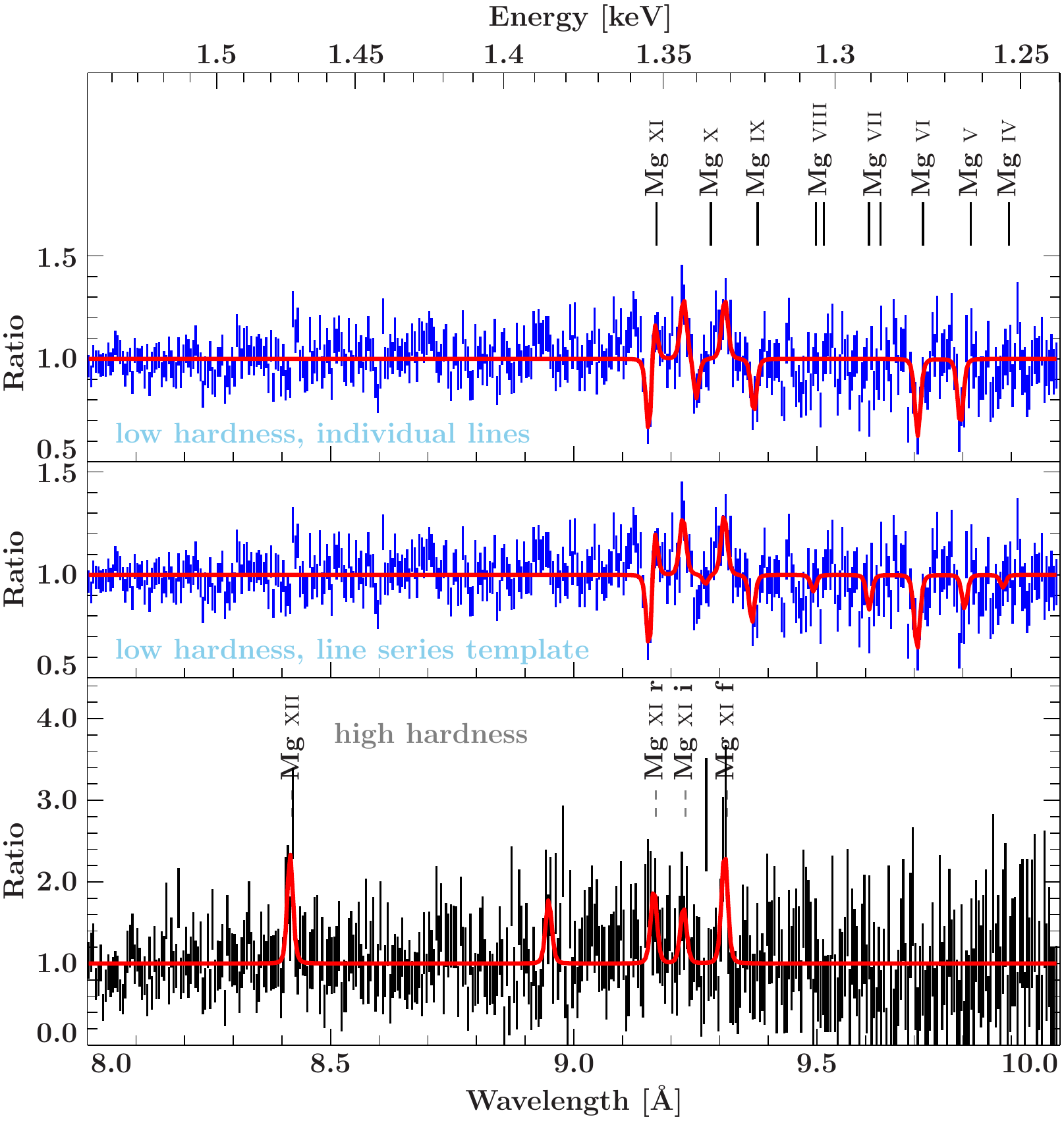}}\hfill
\caption{Magnesium region in low (blue, upper two panels) and high
  (black, lower panel) spectra, corresponding to high and low
  absorption, respectively. The panels show combined HEG and MEG
  $\pm$1 order spectra normalized to their respective
  power-law continua. The solid red line shows the best model
  fit to the data. Two models for the high absorption data are shown
  that differ in their treatment of absorption lines: five individual
  lines with independent Doppler shifts (upper panel) and a template
  consisting of Mg~\textsc{xi} to \textsc{iv} K$\alpha$, all with the
  same Doppler shift (lower panels). We further show the reference
  positions of Mg~\textsc{xi} to \textsc{iv} K$\alpha$
  \citep{Behar_2002a} as well as for the H-like
  \ion{Mg}{xii} Ly$\alpha$ and He-like \ion{Mg}{xi} triplet lines.}\label{fig:Mgratios}
\end{figure}

\begin{figure*}
\begin{minipage}[t]{0.49\textwidth}
\includegraphics[width=\textwidth]{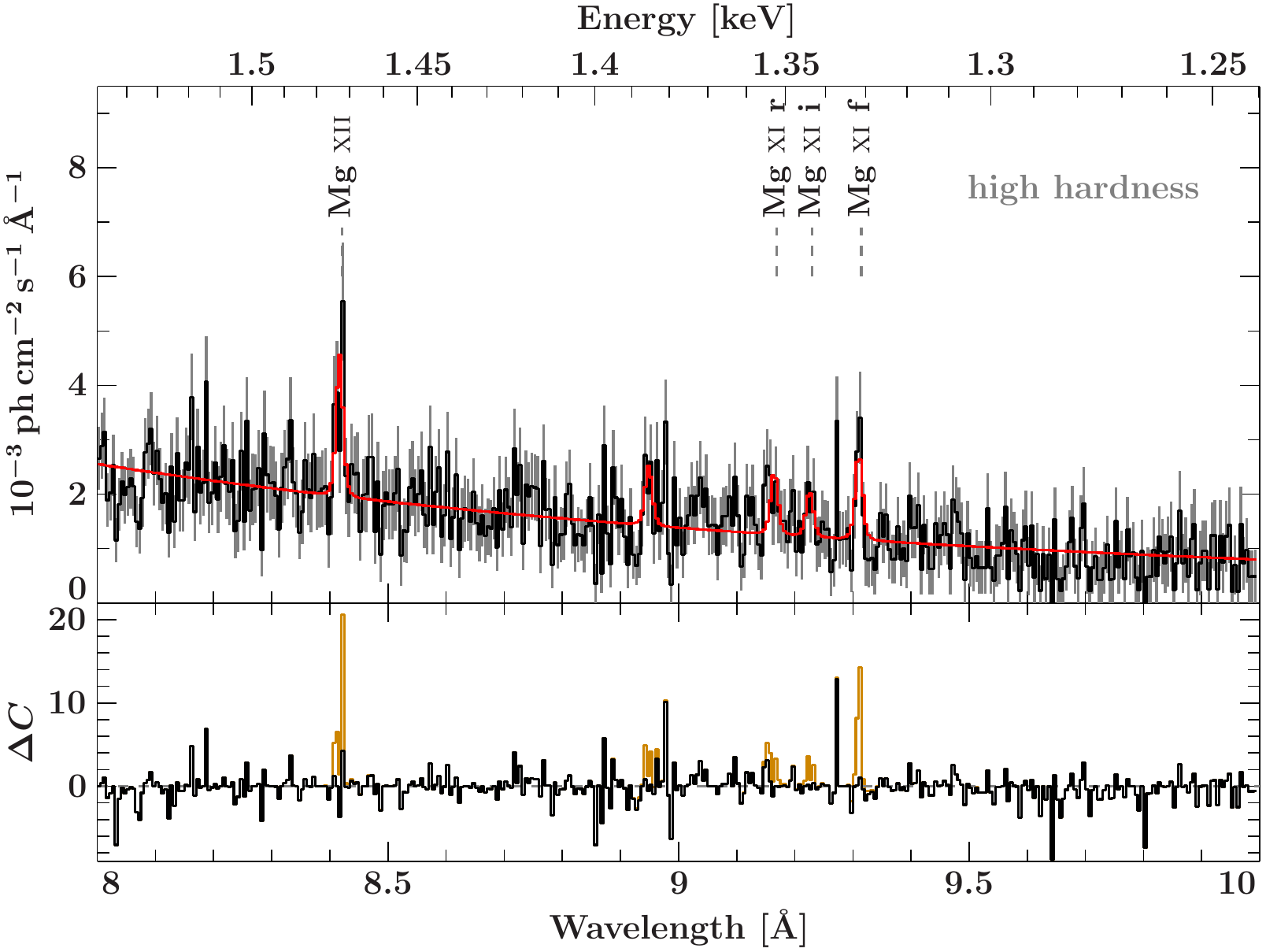}\hfill
\end{minipage}\hfill
\begin{minipage}[t]{0.49\textwidth}
\includegraphics[width=\textwidth]{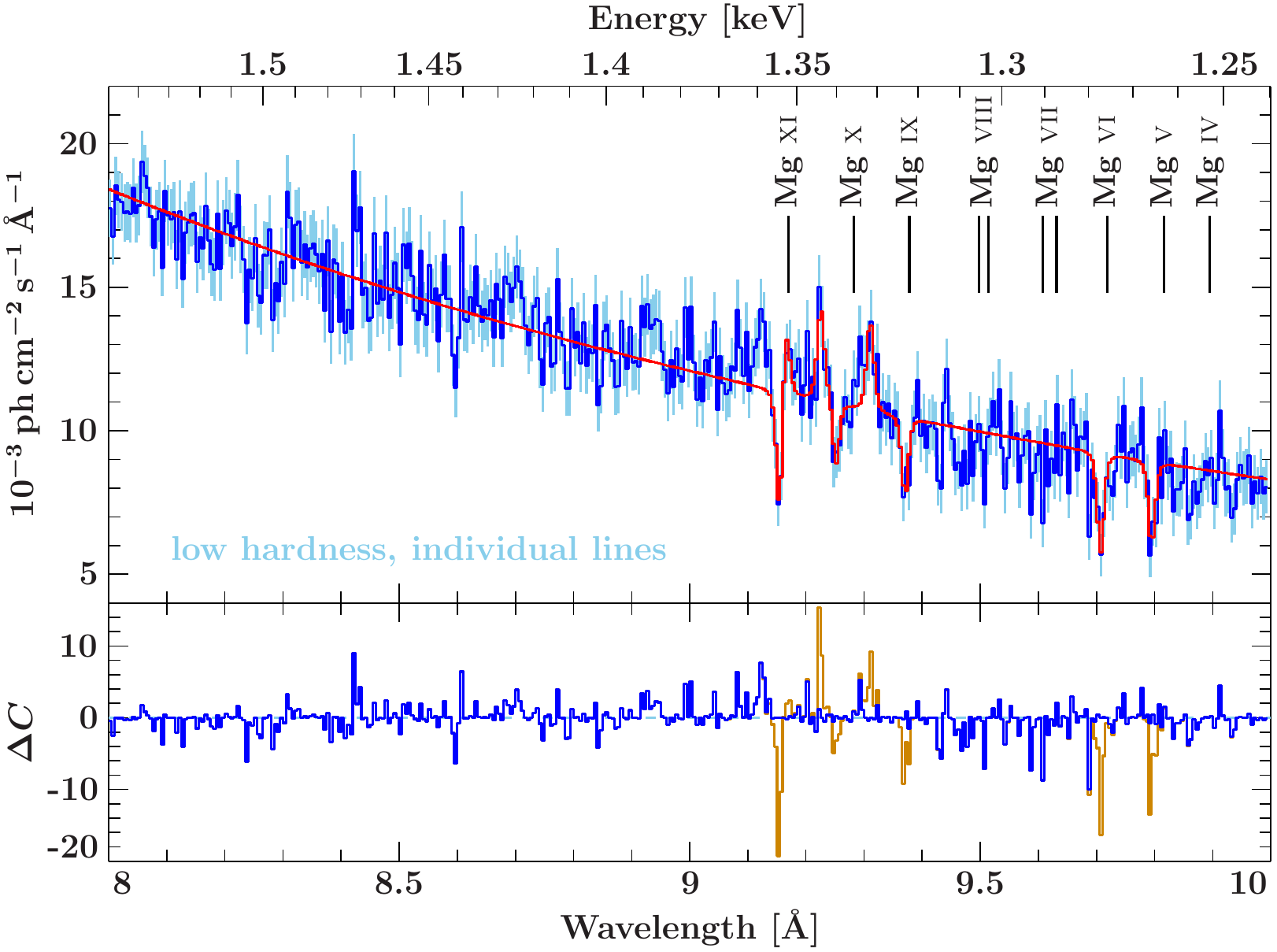}

\includegraphics[width=\textwidth]{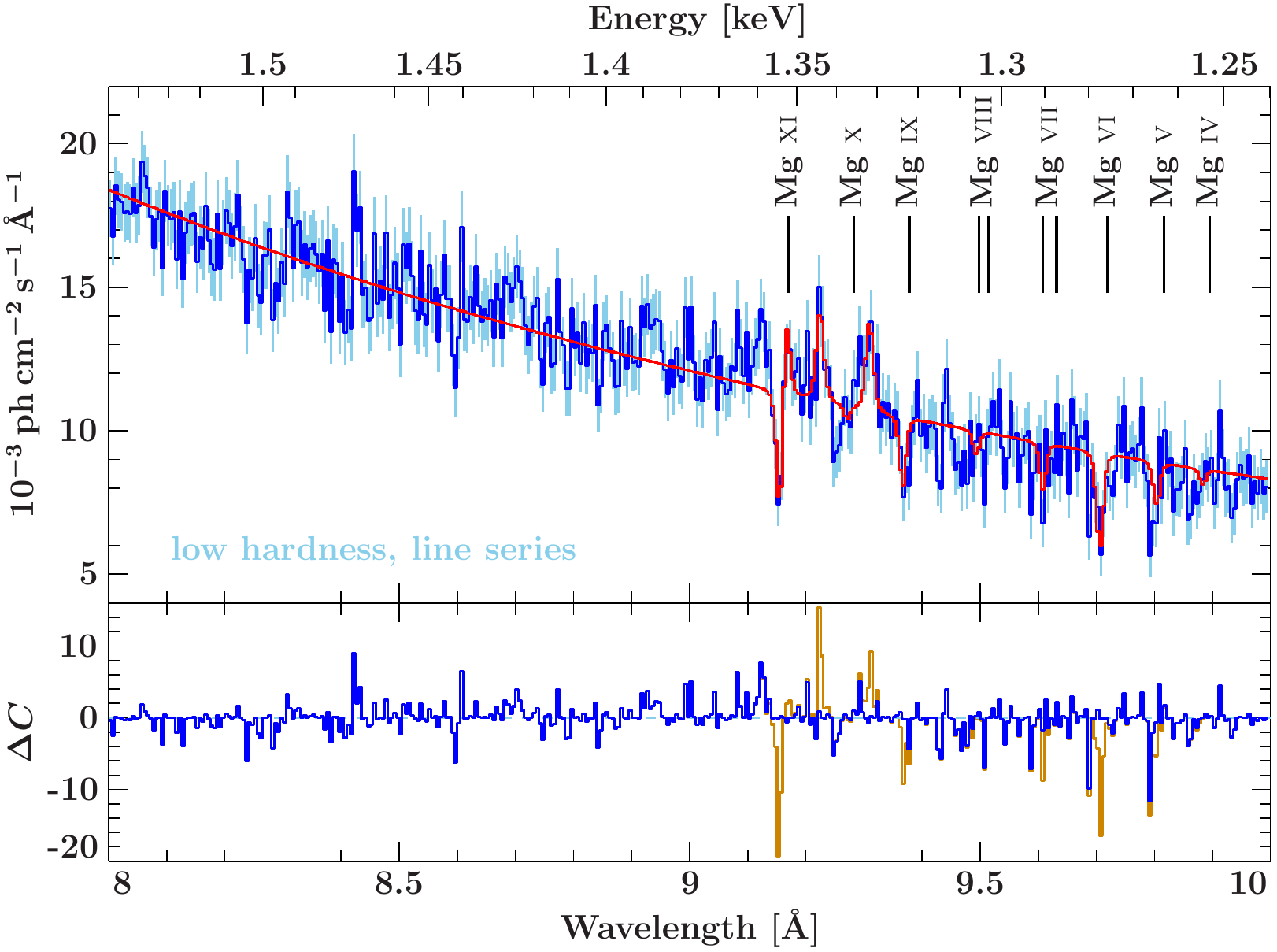}
\end{minipage}
\caption{Mg-region spectra of high hardness (left panel, black) and
  low hardness (right panels, blue) periods. Residuals to best
    fit model are shown in black/blue, residuals to a power law
  without the line components in orange. The upper right panels
  show a fit with five independent absorption lines. The
  lower right panel a fit using a template for Mg-series from
  Mg~\textsc{xi} to \textsc{iv} (reference energies according to
  \citealt{Behar_2002a} indicated by solid lines) with one Doppler
  shift for all lines.  Spectra normalized to their respective
  power-law continua are shown in
  Fig.~\ref{fig:Mgratios}. For the high hardness region lines
    identified are indicated and labelled. For the low hardness region
    we refer to the Tables~\ref{tab:Mgem} and~\ref{tab:Mglow} for the
    identified lines.}
\label{fig:Mgspec}
\end{figure*}

The next detected component consists of a broad excess spanning the
whole He-like \ion{Mg}{xi} triplet (r:
$1\mathrm{s}2\mathrm{p}\,{}^1\mathrm{P}_1$--$1\mathrm{s}^2\,{}^1\mathrm{S}_0$;
i:
$1\mathrm{s}2\mathrm{p}\,{}^3\mathrm{P}_1$--$1\mathrm{s}^2\,{}^1\mathrm{S}_0$
and
$1\mathrm{s}2\mathrm{p}\,{}^3\mathrm{P}_2$--$1\mathrm{s}^2\,{}^1\mathrm{S}_0$,
also known as lines x and y in the notation of
\citealt{Gabriel_1972a}; f:
$1\mathrm{s}2\mathrm{s}\,{}^3\mathrm{S}_1$--$1\mathrm{s}^2\,{}^1\mathrm{S}_0$)
and with a stronger narrower excess at the \ion{Mg}{xi} forbidden
line. We include all three He-line triplet lines in our model, tying
the distance between the triplet lines and the Ly$\alpha$ line to
reference values from \citet[][for He-like]{Drake_1988a} and
\citet[][for H-like]{Erickson_1977a}. The last feature detected with
the Bayesian blocks approach is an emission feature at
$\sim$8.95\,{\AA} ($\alpha_{\mathrm{sig}} = 2$) that we cannot
identify with an obvious line of magnesium or iron
(Figs.~\ref{fig:Mgratios}, lower panel, and \ref{fig:Mgspec}, left
panel).

\begin{table*}
\renewcommand{\arraystretch}{1.3}
\centering 
\caption{Magnesium emission lines in the low- and high-hardness spectra}\label{tab:Mgem}
\begin{tabular}{llcccccc}
  \hline \hline 
  & & reference &Bayesian & $v$ & line &  value\\
  & & wavelength & blocks $\alpha_{\mathrm{sig}}$ & & flux & \\
  & & [\AA]&&  $[$\,km\,s$^{-1}$$]$  & [ph\,s$^{-1}$\,cm$^{-2} \times10^{-5}$] \\
  \hline
  high hardness & \ion{Mg}{xii}~Ly$\alpha$& $8.42101\pm4.0 \times
                                            10^{-5}$ ~\tablefootmark{d}& 5.4 & $140\pm80$ &
                                                               $4.2^{+1.9}_{-1.7}$
                           & --\\
  &  \ion{Mg}{xi}~i & 9.22990\tablefootmark{e}& 5.4\tablefootmark{b} & $ = v_{\ion{Mg}{xii}~\mathrm{Ly}\alpha} $
                    &  $1.8^{+0.5}_{-0.7}$ & --\\
 & $R$ ratio\tablefootmark{c} & -- & -- & -- & -- & $1.5\pm0.7$\,\tablefootmark{f} \\
 & $G$ ratio\tablefootmark{c} & -- & -- & -- & -- & $2.0^{+3.5}_{-0.7}$\,\tablefootmark{f}\\

  \cline{2-7} 
  low hardness &\ion{Mg}{xii}~Ly$\alpha$& $8.42101\pm4.0 \times
                                          10^{-5}$~\tablefootmark{d}& no\tablefootmark{a} & \ldots & \ldots & --\\
                  
  & \ion{Mg}{xi}~r & 9.16896\tablefootmark{e} & $>7$\tablefootmark{b} & $-130\pm60$&
                                                            $4.9^{+2.0}_{-1.9}$ & --\\
  & \ion{Mg}{xi}~i & 9.22990\tablefootmark{e} & $>7$\tablefootmark{b} & $=v_{\ion{Mg}{xi}~\mathrm{r}}$ &
                                                             $5.6^{+2.1}_{-2.0}$& --\\
& \ion{Mg}{xi}~f & 9.31455 \tablefootmark{e} &  $>7$\tablefootmark{b}& $=v_{\ion{Mg}{xi}~\mathrm{r}}$&
                                                          $5.4^{+2.2}_{-2.1}$& --\\
  \hline

\end{tabular}

\tablefoot{
\tablefoottext{a}{We run line searches for $\alpha_{\mathrm{sig}} \geq 1.5$.
}
\tablefoottext{b}{The He-like triplet is detected as a single
  excess feature covering all three lines (r, f and i).
}
\tablefoottext{c}{All three triplet lines are assumed to have the same velocity.}
\tablefoottext{d}{\citet{Erickson_1977a}}
\tablefoottext{e}{\citet{Drake_1988a}; for a discussion of errors on
  \citet{Drake_1988a} values, see Notes to Table~\ref{tab:Sihard}.}
\tablefoottext{f}{These ratios formally correspond to fluxes
    of $2.7 \times10^{-5}$\,ph\,s$^{-1}$\,cm$^{-2}
                                $ for \ion{Mg}{xi}~f and $ 2.1 \times10^{-5}$\,ph\,s$^{-1}$\,cm$^{-2}
                                $ for \ion{Mg}{xi}~r.}
}
\end{table*}

We parametrize the line strengths of the He-like triplet using the
ratios $R = f / i$ and $G = (f + i) / r$
\citep{Gabriel_1969a,Porquet_2000a} and model these density and
temperature diagnostics directly. We obtain $R =1.5\pm0.7$ and
$G = 2.0^{+3.5}_{-0.7}$ and a line shift of $-140\pm80$\,km\,s$^{-1}$
for the identified H- and He-like Mg lines (see Table~\ref{tab:Mgem}
for detailed line parameters of the identified lines). 

It has been recently shown by \citet{Mehdipour_2015a} that, under astrophysical conditions, He-like triplet lines of a given element can be affected by absorption from Li-like ions in the same medium. While \citet{Mehdipour_2015a} discuss this effect for the cases of iron and oxygen, it can easily be extended to other elements. Contamination of the He-like magnesium triplet in emission by absorption lines is clearly seen in the following only for the low-hardness spectrum (Sec.~\ref{sec:Mglow}). While we do not detect any absorption lines in the high-hardness spectrum, we cannot exclude some contamination of the triplet.

\subsubsection{Low-hardness spectrum of the magnesium region}\label{sec:Mglow}

The Bayesian blocks approach finds discrete absorption features at
$\sim$9.15\,{\AA} ($\alpha_{\mathrm{sig}} >$ 7), $\sim$9.29\,{\AA}
($\alpha_{\mathrm{sig}} >$ 4.6), $\sim$9.7\,{\AA} and $\sim$9.8\,{\AA}
($\alpha_{\mathrm{sig}} =$ 2.0). It also finds a wide excess at
$\sim$9.1--9.3\,{\AA}, consistent with the location of the
Mg~\textsc{xi}~triplet that becomes more pronounced as the narrow
absorption features are iteratively taken into account by the
model. We model the He-like triplet as in the case of the high
absorption spectrum by fixing the 
relative location of the lines but leaving one line energy and the
strengths of the individual lines free.  The last component is a
discrete absorption feature at $\sim$9.77\,{\AA}
($\alpha_{\mathrm{sig}} = 2.4$).

The resulting model includes 8 lines: the He-like triplet lines in
emission and five absorption lines (Figs.~\ref{fig:Mgratios}, upper
panels, and \ref{fig:Mgspec}, right upper panel). In spite of suggestive
residuals, no significant \ion{Mg}{xii}~Ly$\alpha$ is found and we
leave it out in our best-fit model.

Since the absorption lines are weak and their widths unresolved, we are
able to replace the narrow Gaussian absorption components used so far
with a simple multiplicative
box-function\footnote{\texttt{narrowline} function in the
  \texttt{isisscripts}
  (\url{http://www.sternwarte.uni-erlangen.de/isis/}); the values for
  the function centered at $\nu_0$ with equivalent width EQW are: 0 in
  $ [\nu_0 - \frac{1}{2}\mathrm{EQW}, \nu_0 + \frac{1}{2}
  \mathrm{EQW}]$ and 1 everywhere else.} and thus fit their
equivalent widths directly.  For the emission lines of the He-like
triplet we obtain velocities of $-130\pm60$\,km\,s$^{-1}$
(He-like triplet line parameters in Table~\ref{tab:Mgem}).

\begin{table*}
\renewcommand{\arraystretch}{1.3}
\centering
\caption{Magnesium absorption lines in the low-hardness spectrum}\label{tab:Mglow}
\begin{tabular}{llccccccc}
\hline \hline
& & & \multicolumn3c{individual lines} &  \multicolumn2c{line series}\\
ion & line & reference\tablefootmark{a}& Bayesian & $v$ & equivalent & $v$ & equivalent\\
 & & wavelength &  blocks $\alpha_{\mathrm{sig}}$ & & width & & width\\
 & & [\AA]& & $[$\,km\,s$^{-1}$$]$ & $[10^{-3}$\,{\AA}$]$ & $[$\,km\,s$^{-1}$$]$ &  $[10^{-3}$\,{\AA}$]$\\
\hline
He-like & \ion{Mg}{xi}~r& 9.170 & $\geq 7$& $-480\pm60$ & $-6.9\pm1.5$ &
                                                                       $-360\pm60$ & $-12.0\pm1.7$\\
Li-like & \ion{Mg}{x} K$\alpha$ & 9.282 & $\geq4.6$ &$-990^{+160}_{-100}$&
                                                       $-3.0\pm1.5$ &$=v_{\ion{Mg}{xi}~r}$& $-0.7\pm2.0$\\
Be-lie & \ion{Mg}{ix} K$\alpha$&9.378&no\tablefootmark{b}&$-250^{+160}_{-130}$ & $-4.1\pm1.6$&$=v_{\ion{Mg}{xi}~r}$&
                                                                      $-3.6\pm1.6$\\
B-like & \ion{Mg}{viii} K$\alpha$&9.619& no\tablefootmark{b}& \ldots & \ldots &$=v_{\ion{Mg}{xi}~r}$& $-1.2\pm1.8$
\\
C-like & \ion{Mg}{vii} K$\alpha$&9.506& no\tablefootmark{b} & \ldots & \ldots &$=v_{\ion{Mg}{xi}~r}$& $-2.6\pm1.9$\\
N-like& \ion{Mg}{vi} K$\alpha$& 9.718 & 2 & $-330\pm90$ &
                                                      $-6.0\pm1.8$
    &$=v_{\ion{Mg}{xi}~r}$& $-5.9\pm1.7$\\
O-like & \ion{Mg}{v} K$\alpha$&9.816 & 2 & $-640\pm90$ & $-5.3\pm1.8$&$=v_{\ion{Mg}{xi}~r}$&
                                                                       $-2.6^{+1.7}_{-2.0}$ \\
F-like & \ion{Mg}{iv} K$\alpha$& 9.895& no\tablefootmark{b}& \ldots & \ldots &$=v_{\ion{Mg}{xi}~r}$ & $-0.9\pm1.8$\\
\hline
\end{tabular}
\tablefoot{
\tablefoottext{a}{All from \citet{Behar_2002a}; values for B-like
  \ion{Mg}{viii}~K$\alpha$ and C-like \ion{Mg}{vii}~K$\alpha$ averages
  over the two strong transitions listed.}
 \tablefoottext{b}{We run line searches for $\alpha \geq 1.5$.}

}
\end{table*}

The locations of the absorption lines are indicative of the different
ionization stages of Mg. We use the theoretical reference values for
the line energies of He-like \ion{Mg}{xi}~r to F-like \ion{Mg}{iv}
K$\alpha$~lines from \citet{Behar_2002a} to identify the lines
(Fig.~\ref{fig:Mgratios}) and calculate the line velocities
(Table~\ref{tab:Mglow}). 

The \ion{Ne}{x}~Ly$\gamma$ line at
9.70818\,{\AA} \citep{Erickson_1977a} 
is very close to N-like \ion{Mg}{vi} K$\alpha$ at 9.718\,{\AA} and
indeed, \citet{Goldstein_2004a} have identified this absorption
feature with \ion{Ne}{x}~Ly$\gamma$. While we detect
\ion{Ne}{x}~Ly$\beta$ (Sec.~\ref{sec:Nelow}), this line is
blue-shifted by $-380^{+150}_{-320}$\,km\,s$^{-1}$; a
\ion{Ne}{x}~Ly$\gamma$ line with a similar shift does not describe the
observed absorption feature. We do also not detect the
neighboring \ion{Ne}{x}~Ly$\delta$ at 9.48075\,{\AA}
\citep{Erickson_1977a}, 
although our upper limit on the equivalent width of this line is
$-4 \times 10^{-3}$\,{\AA}, i.e., within the margins of our
expectations of the strength of \ion{Ne}{x}~Ly$\delta$ given the
equivalent width of \ion{Ne}{x}~Ly$\beta$. On the other hand,
assuming, as done here, that the two prominent absorption features at
9.7--9.8\,{\AA} are from K$\alpha$ transitions in N-like \ion{Mg}{vi} and
O-like \ion{Mg}{v} results in similar strengths for both
lines, which
both appear blue-shifted, consistent with the detected higher
ionization Mg ions, confirming our line identification. We cannot,
however, exclude a contribution of \ion{Ne}{x}~Ly$\gamma$ to the
observed feature we identify as \ion{Mg}{vi}~K$\alpha$ in
absorption. To test the amount of this contribution, we model the
observed absorption feature with two components, one of them at the
expected location of blue-shifted \ion{Ne}{x}~Ly$\gamma$. We obtain
similar velocity and equivalent width for \ion{Mg}{vi}~K$\alpha$ as
without the neon contribution ($v =
-280\pm90$\,km\,s$^{-1}$ and $-5.1^{+1.8}_{-1.6}\times10^{-3}$\,{\AA})
and an equivalent width of $-2.4^{+2.2}_{-1.6}\times10^{-3}$\,{\AA}
for  \ion{Ne}{x}~Ly$\gamma$.

The obtained velocity values for the absorption lines of Mg span a
wide range of $-$(300--900)\,km\,s$^{-1}$ (Table~\ref{tab:Mglow}) and
are inconsistent with each other within their uncertainties. However,
the complexity of the spectrum, in particular the overlap of emission
and absorption features can lead to modelling problems that the formal
errors do not account for. Additionally, inaccuracies in theoretical
reference values may be present: while no EBIT laboratory data are yet
available for magnesium, \citet{Hell_2016a} have shown that for
silicon and sulfur laboratory values and theoretical calculations of
\citet{Behar_2002a} can disagree at up to 300\,km\,s$^{-1}$,
especially for lower ionization stages.

To better assess the contributions of the Mg absorption lines, we
build a template (\ion{Mg}{xi} to \ion{Mg}{iv} K$\alpha$,
Fig.~\ref{fig:Mgratios} and Table~\ref{tab:Mglow}) using the \citet{Behar_2002a} values: we
assume the same line shift for all lines, but let the line strengths
vary.  For B-like \ion{Mg}{viii} and C-like \ion{Mg}{vii},
\citet{Behar_2002a} list two strong transitions. For each ion,
both transitions are of comparable oscillator strength but are,
respectively, 0.016\,{\AA} and 0.024\,{\AA} apart, which corresponds
to a velocity difference of $\sim$500\,km\,s$^{-1}$ and
$\sim$750\,km\,s$^{-1}$, respectively (cf. Fig.~\ref{fig:Mgratios}
where the line positions are marked). We incorporate these lines in
our template using one line with the average line center as listed in
Table~\ref{tab:Mglow}, but note that this can lead to additional
uncertainties in the determination of the line Doppler shifts. We also
include the He-like \ion{Mg}{xi} triplet in emission with parameters
as described above and show the resulting best-fit model in
Figs.~\ref{fig:Mgratios}, middle panel, and \ref{fig:Mgspec}, lower
right panel. The line template parameters are listed in
Table~\ref{tab:Mglow}. The template line velocity of
$-360\pm40$\,km\,s$^{-1}$ is in rough agreement with the average
velocity of the five independent absorption lines. The equivalent
widths of the template lines show that B-like \ion{Mg}{viii} and
F-like \ion{Mg}{iv} K$\alpha$ are consistent with 0, i.e., the lines
are not significantly detected, consistent with their non-detection in
the Bayesian blocks approach. The equivalent width of C-like
\ion{Mg}{vii} K$\alpha$ may indicate a weak line. The large
uncertainties on Li-like \ion{Mg}{x} K$\alpha$, on the other hand, are
due to the blending of the line with the intercombination and
forbidden emission lines of the He-like \ion{Mg}{xi} triplet.

\subsection{Neon region}\label{sec:Ne}

The region of interest encompasses the wavelengths 10--14\,{\AA},
i.e., from H-like \ion{Ne}{x}~Ly$\beta$ to the He-like
\ion{Ne}{ix} triplet. At higher wavelengths, low source counts preclude
an analysis and in particular the detection of possible K$\alpha$
features of lower ionization stages than \ion{Ne}{viii}. All
  identified neon lines in both low- and high-hardness spectra are
  listed in Table~\ref{tab:Ne} together with their respective
  reference values.

\begin{figure*}
\includegraphics[width=0.49\textwidth]{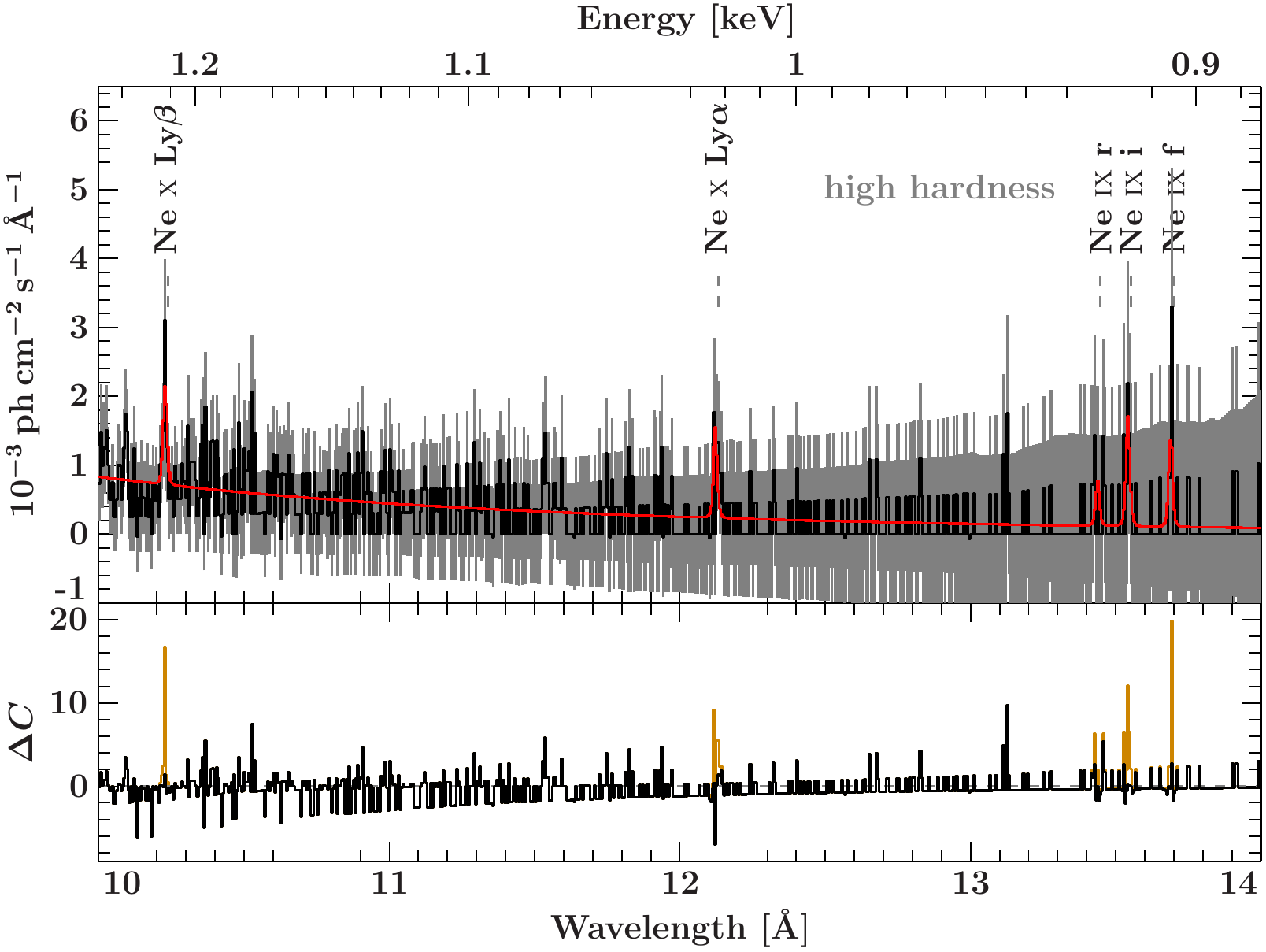}\hfill
\includegraphics[width=0.49\textwidth]{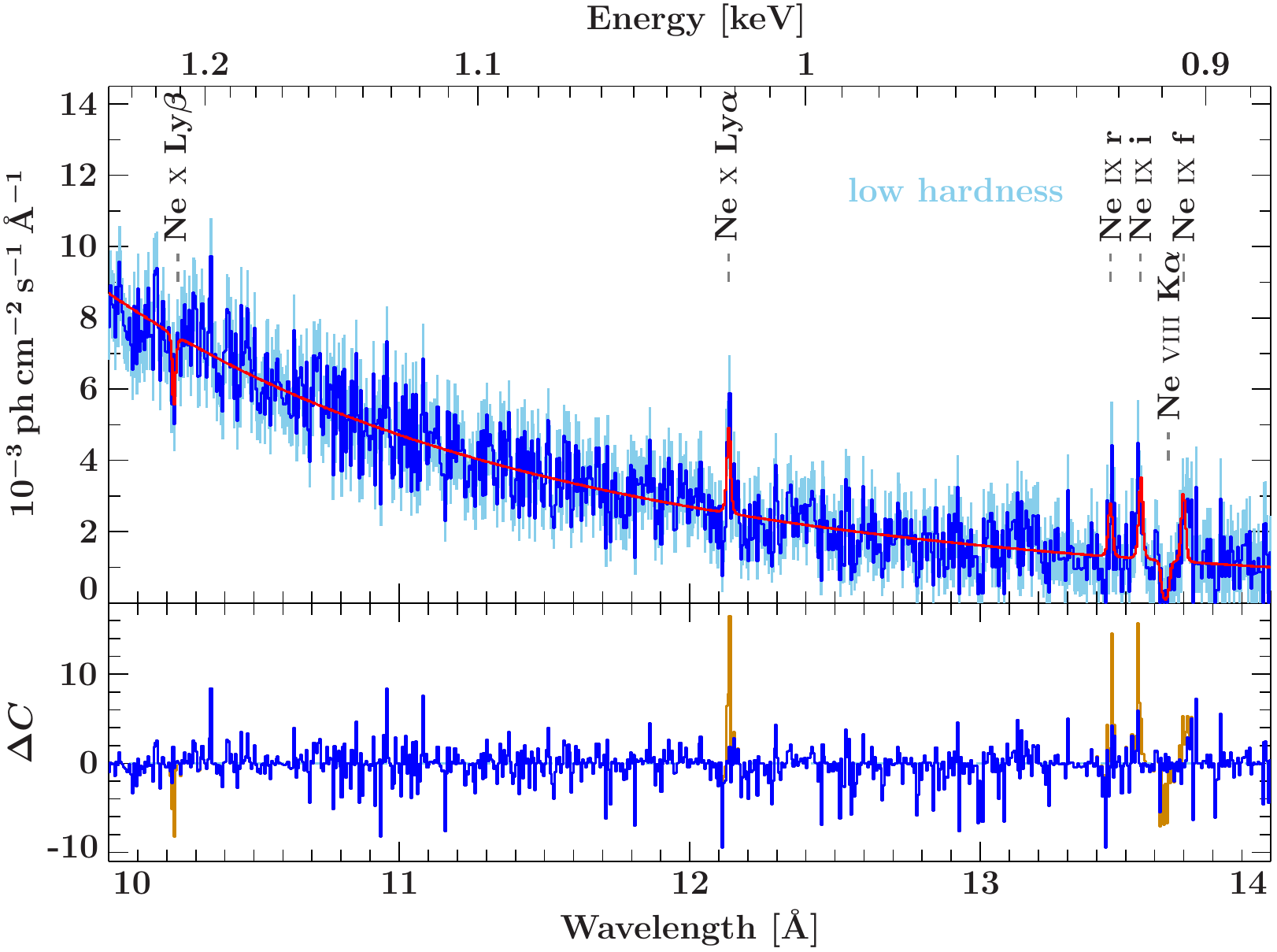}
\caption{Neon region spectra. \textsl{Left panel} (black):
  high-hardness spectrum. \textsl{Right panel} (blue):
  low-hardness spectrum. Best-fit models shown in
  red. Residuals of the Cash statistics for best fit including the
  lines shown in black and blue, respectively; residuals to the power
  law only in orange. The identified lines are indicated and
    labelled.}\label{fig:ne}
\end{figure*}

\subsubsection{High-hardness spectrum of the neon region}\label{sec:Nehigh}

The Bayesian blocks approach finds H-like \ion{Ne}{x}~Ly$\alpha$ with $\alpha_{\mathrm{sig}} = 3.4$, an excess spanning the location of the He-like \ion{Ne}{ix} triplet with $\alpha_{\mathrm{sig}} = 2.5$, and finally the H-like \ion{Ne}{x}~Ly$\beta$ with $\alpha_{\mathrm{sig}} = 2$.  No further structure is detected with this approach once these five lines are included.

We fit just one line shift for all five lines. Similarly to Sec.~\ref{sec:Mghigh}, we directly fit the He-like triplet line ratios, $R$ and $G$ \citep[e.g.,][]{Porquet_2000a} and obtain $R = 0.8^{+1.4}_{-0.6}$ and $G = 4.3^{+5.7}_{-2.9}$. The resulting line velocity is $-220\pm70$\,km\,s$^{-1}$ (Fig.~\ref{fig:ne}).

\begin{table*}
\renewcommand{\arraystretch}{1.3}
\centering 
\caption{Neon lines in the low- and high-hardness spectra}\label{tab:Ne}
\begin{tabular}{llcccccc}
  \hline \hline 
  & & reference &Bayesian & $v$ & line & equivalent & value\\
  & & wavelength & blocks $\alpha_{\mathrm{sig}}$ & & flux &width & \\
  & & [\AA]& &  $[$\,km\,s$^{-1}$$]$  & [ph\,s$^{-1}$\,cm$^{-2}
                                \times10^{-5}$] & [\AA] \\
  \hline
  high hardness & \ion{Ne}{x}~Ly$\alpha$ & 12.1338\tablefootmark{d} & 3.4 & $-220\pm70$ &
                                                               $2.1^{+1.7}_{-1.2}$
                           & -- & --\\
 & \ion{Ne}{x}~Ly$\beta$ & 10.23887\tablefootmark{d} & 2 & $=v_{\ion{Ne}{x}~\mathrm{Ly}\alpha}$ &
                                                                      $2.5^{+1.5}_{-1.2}$
                           & -- & --\\
  &  \ion{Ne}{ix}~i & 13.5520\tablefootmark{e} &2.5 & $=v_{\ion{Ne}{x}~\mathrm{Ly}\alpha}$ &
                                                                  $3.0^{+2.3}_{-1.6}$
                           & -- & --\\
 & $R$ ratio\tablefootmark{c} & -- & -- & -- & -- & -- &$0.8^{+1.4}_{-0.6}$\,\tablefootmark{g}  \\
 & $G$ ratio\tablefootmark{c} & --& -- & -- & -- & -- & $4.3^{+5.7}_{-2.9}$\,\tablefootmark{g}\\

  \cline{2-8} 
  low hardness &\ion{Ne}{x}~Ly$\alpha$ & 12.1338\tablefootmark{d} & 4.6 & $40^{+80}_{-50}$ &
                                                                  $4.3^{+1.9}_{-1.7}$
                           & -- & --\\
                  
  &\ion{Ne}{x}~Ly$\beta$ & 10.23887\tablefootmark{d} & no\tablefootmark{a} & $-380^{+150}_{-320}$
                    & --&
                          $\left(-4.3^{+2.2}_{-2.0}\right)\times10^{-3}$
                                        & --\\
                  
  & \ion{Ne}{ix}~r & 13.4476\tablefootmark{e} & 5\tablefootmark{b} &
                                          $=v_{\ion{Ne}{x}~\mathrm{Ly}\alpha}$&
                                                                                $2.8^{+1.7}_{-1.8}$
                           & -- & --\\
  & \ion{Ne}{ix}~i & 13.5520\tablefootmark{e} & 5\tablefootmark{b}&
                                         $=v_{\ion{Ne}{x}~\mathrm{Ly}\alpha}$
                    & $4.3^{+2.6}_{-1.6}$ & -- & --\\
& \ion{Ne}{ix}~f & 13.6993\tablefootmark{e}& 5\tablefootmark{b}&
                                       $=v_{\ion{Ne}{x}~\mathrm{Ly}\alpha}$&
                                                                             $3.7^{+2.0}_{-2.1}$
                           & -- & -- \\
& \ion{Ne}{viii}~K$\alpha$ & 13.647\tablefootmark{f} & 5\tablefootmark{b} &  $-220\pm140$ & -- &
                                                                       $-0.029^{+0.011}_{-0.009}$
                                        & --\\
  \hline

\end{tabular}
\tablefoot{
\tablefoottext{a}{We run line searches for $\alpha \geq 1.5$.
}
\tablefoottext{b}{The He-like triplet emission lines and the Li-like
  absorption line are detected as one complex-shaped feature.
}
\tablefoottext{c}{All three triplet lines are assumed to have the same velocity.
}
\tablefoottext{d}{\citet{Erickson_1977a}}
\tablefoottext{e}{\citet{Drake_1988a}; for a discussion of errors on
  \citet{Drake_1988a} values, see Notes to Table~\ref{tab:Sihard}.}
\tablefoottext{f}{\citet{Behar_2002a}}
\tablefoottext{g}{These ratios formally correspond to fluxes
    of $2.4 \times10^{-5}$\,ph\,s$^{-1}$\,cm$^{-2}
                                $ for \ion{Ne}{ix}~f and $0.7 \times10^{-5}$\,ph\,s$^{-1}$\,cm$^{-2}
                                $ for \ion{Ne}{ix}~r.  }
}
\end{table*}

\subsubsection{Low-hardness spectrum of the neon region}\label{sec:Nelow}

The Bayesian Blocks approach finds a complex feature that we identify
with the He-like \ion{Ne}{ix} triplet lines ($\alpha_{\mathrm{sig}} = 5$) and model
with Gaussian components. The feature shows a dip between the
\ion{Ne}{ix}~f and \ion{Ne}{ix}~i component and we model it with a
\texttt{narrowline} component that we identify with the Li-like
\ion{Ne}{viii} K$\alpha$. 
The next strongest feature is H-like \ion{Ne}{x} at
$\alpha_{\mathrm{sig}} = 4.6$ that we fit with a Gaussian. We also
note that the strong \ion{Fe}{xvii}
$1s^2\,2s^2\,2p_{1/2}\,2p_{3/2}^4\,4d_{3/2}
{}^1P_1$--$1s^2\,2s^2\,2p^6 {}^1S_0$ transition (commonly
  denoted as 4C) is located at 12.124\,{\AA} \citep{Brown_1998a},
i.e. close to H-like \ion{Ne}{x}, but we do not detect any other
strong lines from iron, making an identification of the observed
absorption feature with this line doubtful.

The presence of the \ion{Ne}{viii} K$\alpha$ line hints towards a complex interplay of
absorption and emission features as also already seen in the
low-hardness spectrum of the magnesium region (Sec.~\ref{sec:Mglow}).

We tie the centers of the four emission features together and obtain a
line velocity of $40^{+80}_{-50}$\,km\,s$^{-1}$. For \ion{Ne}{viii} K$\alpha$, we
obtain $v = -220\pm140$\,km\,s$^{-1}$, but given the low counts in the
spectrum and the interplay of the emission and absorption, the formal
errors are likely underestimated.

Because of the presence of H-like \ion{Ne}{x}~Ly$\beta$ in the
low-hardness spectrum and suggestive residuals at
$\sim$10.2\,{\AA}, we add this line in absorption to our model and
obtain an equivalent width of $-0.029^{+0.011}_{-0.009}$\,{\AA} and a
line velocity of $-380^{+150}_{-320}$\,km\,s$^{-1}$. This velocity is
inconsistent with that of the emission lines and attempts to tie the
velocities do not result in a satisfactory fit. The presence of
\ion{Ne}{x}~Ly$\beta$ does not change the modelling results for other
lines in the region.

For a single uniform medium, lines of the same series are
expected to be all either in absorption or in emission. However, in
the case of Vela X-1 we expect a complex geometry, as discussed in
Sec.~\ref{sec:discussion}, and the sum of emission and absorption line
contributions from different regions could result in the observed
properties of the \ion{Ne}{x}~Ly series. The overall behavior however
remains puzzling.

\subsection{Iron K$\alpha$ region}

\begin{figure*}
\includegraphics[width=0.49\textwidth]{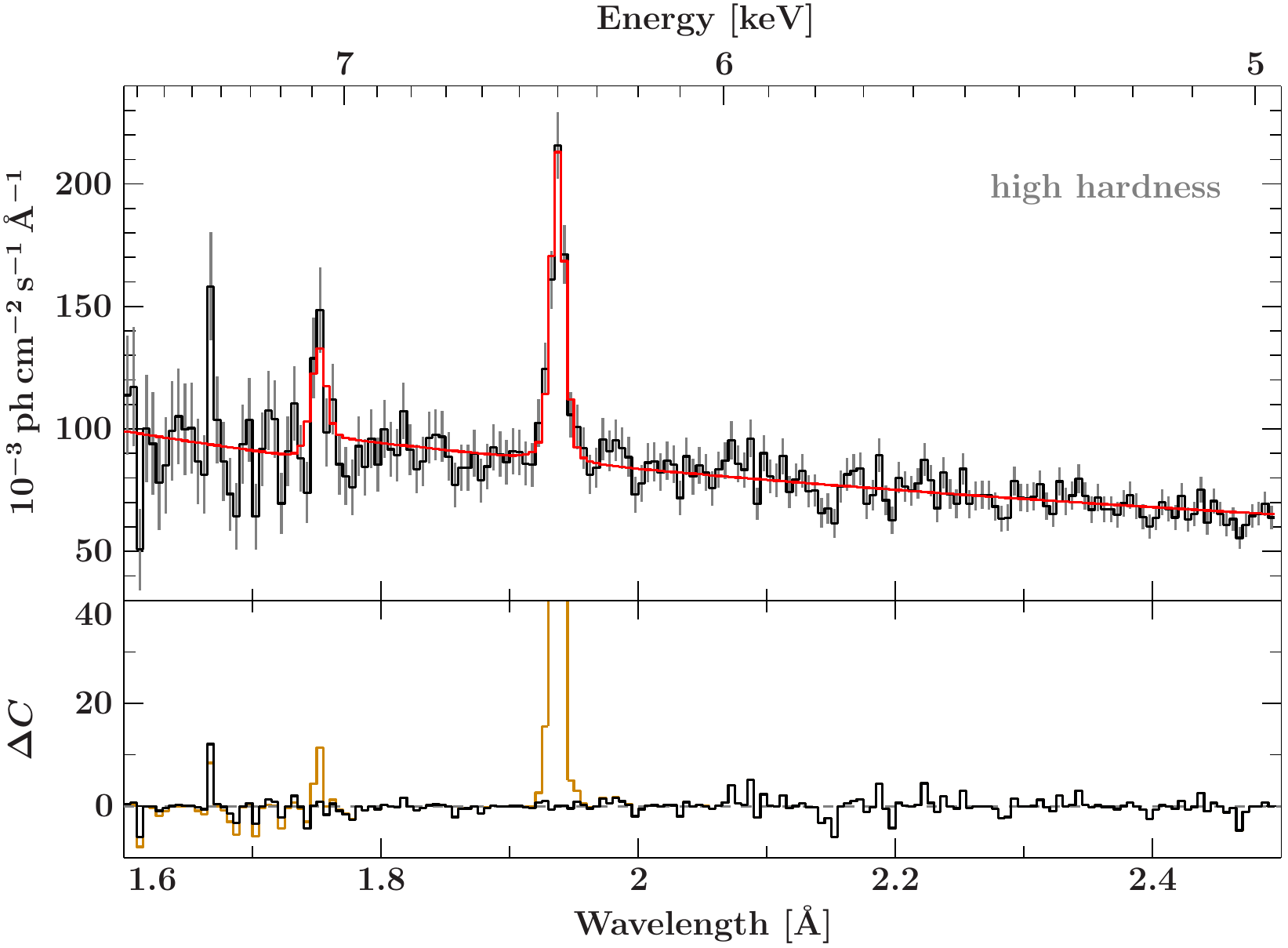}\hfill
  \includegraphics[width=0.49\textwidth]{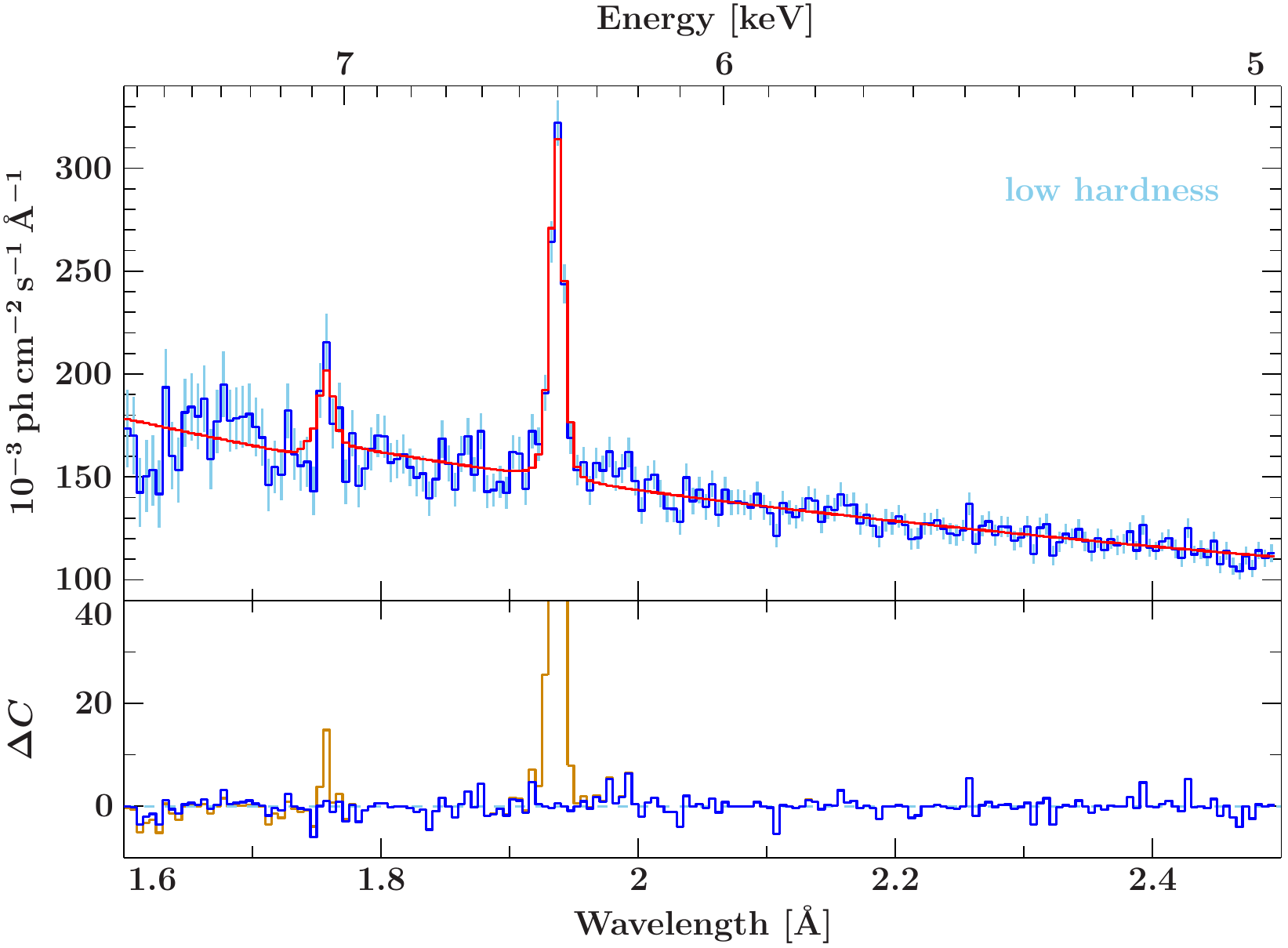} \caption{Fe
  region spectra. \textsl{Left panel}: high-hardness
  spectrum. \textsl{Right panel}: low-hardness spectrum. Data
  are shown in upper panels in black and blue, best-fit model in red.
  Residuals of best fit including the lines and edge
  (high-hardness
  spectrum) are shown in black and blue, respectively; residuals for the
  power law only are shown in orange. }\label{fig:Fe}
\end{figure*}

The region encompasses wavelengths between 1.6--2.5\,{\AA}, i.e., the
locations of the neutral Fe~K$\alpha$ (1.937\,{\AA}) and Fe~K$\beta$
(1.757\,{\AA}) lines as well as the iron edge (1.740\,{\AA}) and
Ni~K$\alpha$ (1.66\,{\AA}) line. The use of a broader region than
covered by the line features themselves allows better constraints on
the power-law continuum. The results for this section are listed in
Table~\ref{tab:Fe} and shown in Fig.~\ref{fig:Fe}.

In the high-hardness spectrum, the Bayesian Blocks approach first
finds the obviously present Fe~K$\alpha$ line, followed by the
Fe~K$\beta$ and a step-like feature that we identify as the iron
edge. We model the lines with Gaussian components and the edge with
the \texttt{edge} model, fixing the edge energy to 1.740\,{\AA}
(Fig.~\ref{fig:Fe}). The resulting line and edge parameters are listed
in Table~\ref{tab:Fe}. Ni~K$\alpha$ has been previously detected in
Vela X-1 \citep[e.g.,][]{Martinez-Nunez_2014a}. However, no
feature at the location of the Ni~K$\alpha$ shortwards in wavelength
from the edge is found with Bayesian Blocks; there is a strong
  nearby residual in the low-hardness spectrum, but given the low
  counts and large uncertainties of this region and the fact that the
  feature is only one bin wide, we do not consider it to be a physical
  Ni~K$\alpha$ line.

\begin{table*}
\renewcommand{\arraystretch}{1.3}
\centering 
\caption{Emission lines and absorption edge in the iron K$\alpha$
  region.}\label{tab:Fe}
\begin{tabular}{llccccc}
  \hline \hline 
  & & Bayesian & $\nu$ & line & equivalent &  $\tau$\\
  & & blocks $\alpha_{\mathrm{sig}}$& & flux & width &\\
  & & & [{\AA}] & [ph\,s$^{-1}$\,cm$^{-2} \times10^{-3}$]& [eV]\tablefootmark{a} \\
  \hline
  high hardness & Fe K$\alpha$ & $\geq 10$ & $1.9375\pm0.0009$ &
                                                     $1.84\pm0.23$
                   & $71\pm9$ & --\\
  &  Fe K$\beta$ & 2.2\tablefootmark{b} & $1.752\pm0.004$ & $0.53\pm0.3$ &
                                                               $23\pm12$
                                & --
  \\
  & Fe K edge & 2.2 \tablefootmark{b} & -- & -- & -- & $0.11 \pm 0.08$ \\
  \cline{2-7} 
  low hardness & Fe K$\alpha$ & $\geq 10$ & $1.9369\pm0.0006$ & $2.43\pm0.18$
                   & $54\pm4$  & --\\
  &  Fe K$\beta$ & 3 & $1.758\pm0.003$ & $0.51\pm0.23$ &  $12\pm6$  & --\\
  & Fe K edge & no & -- & -- & -- & $0.050^{+0.030}_{-0.035}$ \\
  \hline
\end{tabular}
\tablefoot{
  \tablefoottext{a}{Given in eV to enable comparison with
    theoretical expectation in Sec.~\ref{sec:ironfluor}.}
  \tablefoottext{b}{Fe~K$\beta$ and Fe~K~edge found as one
    complex-shaped feature.}
}
\end{table*}

In the low hardness spectrum, the Bayesian Blocks approach finds only 
the Fe~K$\alpha$ and Fe~K$\beta$ lines.  The inclusion of the edge, 
however, still improves the fit and we obtain an absorption edge depth 
of $0.05\pm0.04$, i.e., marginally inconsistent with zero. 

\section{Discussion}\label{sec:discussion}

\subsection{The source of varying absorption column density}\label{sec:discabs}

Past analyses have established a relatively stable behavior of the
central X-ray engine in Vela~X-1 \citep{Odaka_2013a} even during major
flares \citep{Martinez-Nunez_2014a}.  Our modelling in
Sect.~\ref{sec:cont} suggests that the rapidly changing absorption is
at least one of the sources of the observed changes of spectral shape
but we cannot exclude changes in the underlying spectral shape as a
further contribution. Physical origin of possible changes in
  spectral slope has been discussed, e.g., by 
\citet{Odaka_2014a}. Inhomogeneous absorption by clumpy matter has
been used in Vela~X-1 as early as by \citet{Nagase_1986a} to explain
the transient soft excess. Such absorption changes are expected either
if the wind that the neutron star is accreting from is clumpy
\citep[review by][and ref. therein]{Martinez-Nunez_2017a} or if the
inner parts of the accretion flow are highly structured
\citep{Manousakis_2011_PhD,Manousakis_2015a}; both effects could be present
simultaneously.

Clumpy winds are expected for O/B stars: line-driven winds of
such stars are subject to instabilities
\citep{Lucy_1970a,Lucy_1980a}. Hydrodynamical simulations show that
perturbations are present, grow rapidly and cause strong shocks so
that dense gas-shells form
\citep{Owocki_1988a,Feldmeier_1997a}. Density, velocity, or
temperature variations compress the gas further and fragment it into
small, overdense clumps embedded in tenuous hot gas
\citep[e.g.,][]{Oskinova_2012a,Sundqvist_2013a}. In the HMXB Cyg X-1,
for example, the clumpy wind paradigm has been used to explain the
absorption dips, long-term variability of the absorption along the
orbit and the changing line-content observed in high-resolution
spectra during dips
\citep{Hanke_2009a,Grinberg_2015a,Miskovicova_2016a}. In Vela X-1,
clumpy winds can reconcile the discrepancies between different
estimates of the mass-loss rate \citep{Sako_1999a}.  Several lines of
evidence also point towards the presence of wind clumping very close
to the photosphere of O/B stars
\citep{Cohen_2011a,Sundqvist_2013a,Torrejon_2015a}, well within the
$\sim$1.7\,$R_\mathrm{HD~77581}$ distance between the neutron star and
HD~77581 in Vela~X-1.

\begin{figure}
\resizebox{\hsize}{!}{\includegraphics{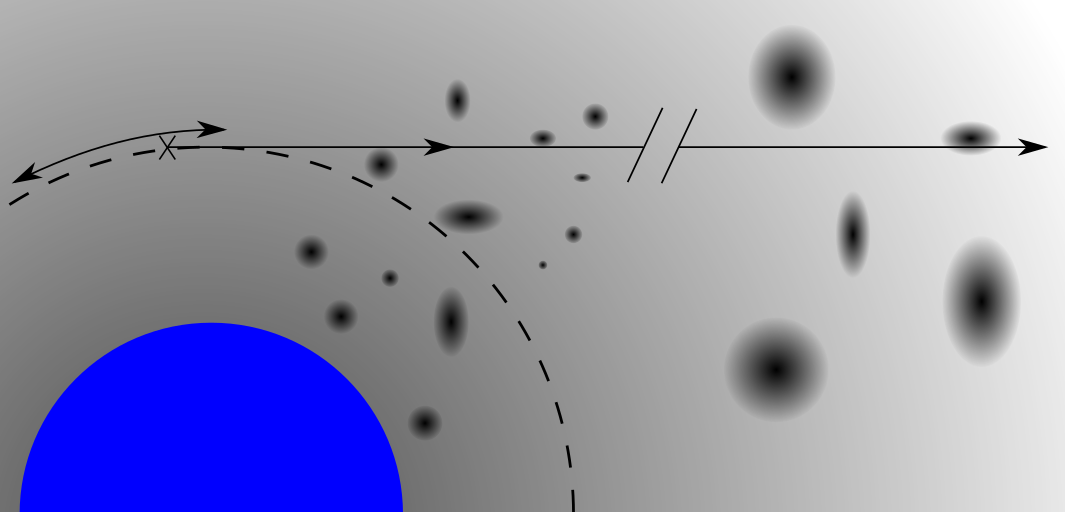}}
\caption{Illustration (not to scale) of the computation in
  Sec.~\ref{sec:discabs}. The inhomogeneous wind induces a
  time-variable column density on the line of sight (solid line) as
  the constant X-ray flux point source (cross) orbits the star (in
  blue) at $\phi_{\mathrm{orb}} = 0.21$--0.25 (arc interval). The
  overdense regions are fiducially represented using radially fading
  spheroidal black dots and the inter-clump environment using a
  smoothly decaying spherical profile.}\label{fig:clumpywind}
\end{figure}

To evaluate the impact of the clumpiness along the line-of-sight on the time variability of the column density, we designed a toy model consisting of a point source with a constant X-ray flux orbiting an O/B star at an orbital separation of 1.8\,$R_\star$ (Fig.~\ref{fig:clumpywind}). The wind is represented relying on the multi-dimensional numerical simulations of \citet{Sundqvist_2017a}. They computed, as a function of time, the mass-density, velocity and pressure profiles in the wind up to 2 stellar radii. Doing so, they witnessed the formation of small overdense regions (clumps), embedded in a lower density environment, and determined their dimensions and mass. We assumed an edge-on inclination and a self-similar expansion of their results to extend their simulation space to up to 15\,$R_\star$ and computed the column density along the line of sight at $\phi_{\mathrm{orb}} = 0.21$--0.25. Beyond the expected systematic decrease in column density as the compact object moves towards us, we observe a 30\% peak-to-peak spread with a correlation time of at most 1ks, a few times longer than the self-crossing time of the clumps. The combined absorption of independent clumps along the line-of-sight thus leads to larger and longer increases in column density than a single clump passing by, but not by a factor of a few in amplitude and not for a few kiloseconds. Therefore, given the characteristic size, speed and distribution of clumps in the wind of isolated massive stars \citep{Sundqvist_2017a}, the inhomogeneities in a wind unperturbed by the presence of the compact object are not enough to account for the observed enhanced absorption (Sec.~\ref{sec:cont}).

However, the compact object can significantly alter the wind structure, either through its gravitational \citep{El_Mellah_2017a} or its radiative influence \citep{Blondin_1990a,Manousakis_2015a}. The so emerging structures (tidal stream, trailing accretion and photoionization wake) highly modulate the systematic orbital evolution of the column density \citep[observational evidence in, e.g.,][]{Kaper_1994a,Fuerst_2010a,Doroshenko_2013a,Malacaria_2016a}.  Yet, their influence should be minimal around $\phi_{\mathrm{orb}} \approx 0.25$ since most of the aforementioned orbital-scale structures lie essentially along the axis joining the two body (star and neutron star) centers or trail the neutron star (Fig.~\ref{fig:sketch}). In this way, they are far enough from the line of sight that we do not expect much of the variation in the column density to be associated to them.  Still, closer to the neutron star, \citet{Manousakis_2015a} showed that even assuming a smooth upstream wind, the flow is highly structured and variable within the shocked region, possibly giving rise to changing absorption such as observed. Further numerical investigations are ongoing and aim at following the clumps as they encounter the shock \citep{El_Mellah_2017b}: the role the clumps play in time-varying absorption would be altered by the deceleration and compression they undergo at the shock. There is also observational support for the presence of such shocks in Vela X-1 since its broadband X-ray count rate has been shown to follow a log-normal distribution \citep{Fuerst_2010a}. This distribution can be described by a multiplicative stochastic model \citep{Uttley_2005a,Fuerst_2010a} and is known to be a possible result of shocks acting on clumpy material \citep[see, e.g.,][]{Kevlahan_2009a}.

Further, newest calculations for stellar-atmosphere models estimate
the wind velocity in the inner region, where the neutron star is
located, to be as low as ${\sim}100\,\mathrm{km}\,\mathrm{s}^{-1}$
\citep{Sander_2017b}, i.e., on the order of the orbital velocity. In such
a case, the wind and accretion structure could be significantly
altered \citep{Shapiro_1976a}, possibly leading to the formation of a
transient disk whose formation and dissipation could, for example,
result in absorption and flaring events. A detailed modelling of such
an accretion flow is outside of the scope of this paper, but we
point out that, to the knowledge of the authors, no clear disk
signatures have been detected in any of the X-ray observations of Vela
X-1 yet.

\subsection{He-like triplet diagnostics: properties of the X-ray emitting gas}

We measured the ratios $R=f/i$ and $G=(f+i)/r$ \citep{Gabriel_1969a,Blumenthal_1972a,Porquet_2000a} for the He-like \ion{Mg}{xi} and \ion{Ne}{ix} triplets in the high-hardness spectra and obtain $R =1.5\pm0.7$ and $G = 2.0^{+3.5}_{-0.7}$ (Sec.~\ref{sec:Mghigh}) and $R = 0.8^{+1.4}_{-0.6}$ and $G = 4.3^{+5.7}_{-2.9}$ (Sec.~\ref{sec:Nehigh}), respectively.  The presence of strong resonance lines (Figs.~\ref{fig:Mgratios},~\ref{fig:ne}) and the measured best fit value of $G < 4$ for magnesium imply that the He-like triplets in the high-hardness spectrum may not be consistent with emission from a purely recombination-dominated photoionized plasma. While we do not use the \ion{Si}{xiii} triplet to calculate $G$ and $R$, the two measured lines in the high-hardness spectrum, \ion{Si}{xiii}~r and \ion{Si}{xiii}~f, are of comparable strengths, while the intercombination line is too weak to be detected (Figs.~\ref{fig:Siratios} and \ref{fig:Sispec} and Table~\ref{tab:Sihard}), supporting the results obtained for Mg and Ne. We do, however, emphasize the large uncertainties of the
  $R$ and $G$ values we obtain from our fits: the following discussion
of plasma diagnostics should thus be seen as a roadmap for future
investigations with missions with a larger collecting area such as
JAXA's X-ray Astronomy Recovery Mission (\textsl{XARM}) or ESA's \textsl{Athena} \citep{Nandra_2013a}.

In principle, $G < 4$ (but note that $G=4$ is still within the
  large measured uncertainties) could be due to the presence of a
collisionally ionized component, which is predicted to have a much
lower $G$ ratio of $\sim$1. However, it has been shown in a number of
other accretion-powered photoionized plasmas that resonance scattering
can also contribute to the emission, with the relative contributions
of resonance scattering and recombination determining the observed $G$
ratio \citep{Sako_2000a, Kinkhabwala_2002a, Kinkhabwala_2003a,
  Wojdowski_2003a}. \citet{Kinkhabwala_2002a} calculated spectra for
different ionic column densities in the ionization cone of a Type~2
Seyfert galaxy (i.e., viewed from the side, with direct emission from
the central source obscured); based on Figure 7 of their article, we
can estimate that an ionic column density of order
$10^{18}\,\mathrm{cm}^2$ is consistent with the observed $G$ ratios of
\ion{Ne}{ix} and \ion{Mg}{xi}.

To estimate the elemental column densities in the wind, we first
calculate the characteristic hydrogen column density of the wind,
$N_\star$, that is defined as \begin{equation}
  N_\star \equiv \frac{\dot{M}}{4\pi \mu m_p R_\star v_\infty}\,
\end{equation} with $\dot M$ the mass-loss rate, $\mu$ the mean weight,
$m_p$ the proton mass, $R_\star$ the
radius of the star and $v_\infty$ the terminal velocity of the wind.
This is the hydrogen column density that would be observed along a
line of sight from the observer to the stellar surface in the case of
constant wind velocity, and is a good order-of-magnitude proxy for the
typical column density experienced by X-rays escaping along different
lines of sight from the neutron star. For the typical values for
HDE~77581, i.e.,
$\dot{M} \sim 2 \times 10^{-6}\,M_\odot\,\mathrm{yr}^{-1}$,
$v_\infty = 750\,\mathrm{km\,s}^{-1}$, and $R_\star = 30\,R_\odot$, we obtain
$N_\star = 2.4 \times 10^{22}\,\mathrm{cm}^{-2}$. Using the abundances of
\citet{Wilms_2000a}, we then estimate typical elemental column
densities of $2 \times 10^{18}$ for \ion{Ne}{ix}, $6 \times 10^{17}$
for \ion{Mg}{xi}, and $4.5 \times 10^{17}$ for \ion{Si}{xiii}. These
crude estimates for the elemental column densities are of order
comparable to the above estimate of the ionic column densities from
the $G$ ratios, implying that a substantial fraction of the material
along the line of sight is ionized to the He-like charge state for
each of these elements. Further, more stringent observational
constrains on $G$ ratio at $\phi_{\mathrm{orb}} \approx 0.25$ will be
crucial to confirm or refute this line of argumentation that can
currently be cast in doubt by the large measured uncertainties on $G$.

The measured values of $R$ are not consistent with expectations for a low-density photoionized plasma. This is particularly striking for neon, where we would expect $R\approx3.1$ for a low-density plasma with $T=2\times10^5$\,K and higher $R$ for higher temperatures \citep{Porquet_2000a}, all values above the measured $R = 0.8^{+1.4}_{-0.6}$.  One possible interpretation is that the plasma density is high; however, the implied densities of $10^{12}$--$10^{13}$\,cm$^{-3}$ \citep[][Fig.~9]{Porquet_2000a} are much higher than expected anywhere in the wind, or even in the accretion wake, and should only be found in the hydrostatic part of the stellar atmosphere. Possible structures close to the neutron star such as discussed in Sec.~\ref{sec:discabs} are unlikely to lead to such extreme density enhancements; typical high densities reached in the simulations of \citet{Manousakis_2011_PhD} or \citet{Mauche_2008a} are of the order of a few 10$^{10}$\,cm$^{-3}$ and below. On the other hand, it is well known from studies of single OB stars \citep{Kahn_2001a} that the strong stellar UV field of HDE~77581 will reduce the forbidden line strength near the star by depopulating the metastable $1\mathrm{s}2\mathrm{s}\,{}^3\mathrm{S}_1$ level \citep{Gabriel_1969a,
  Blumenthal_1972a, Mewe_1978a,Porquet_2001a}.

\begin{figure}
\resizebox{\hsize}{!}{\includegraphics{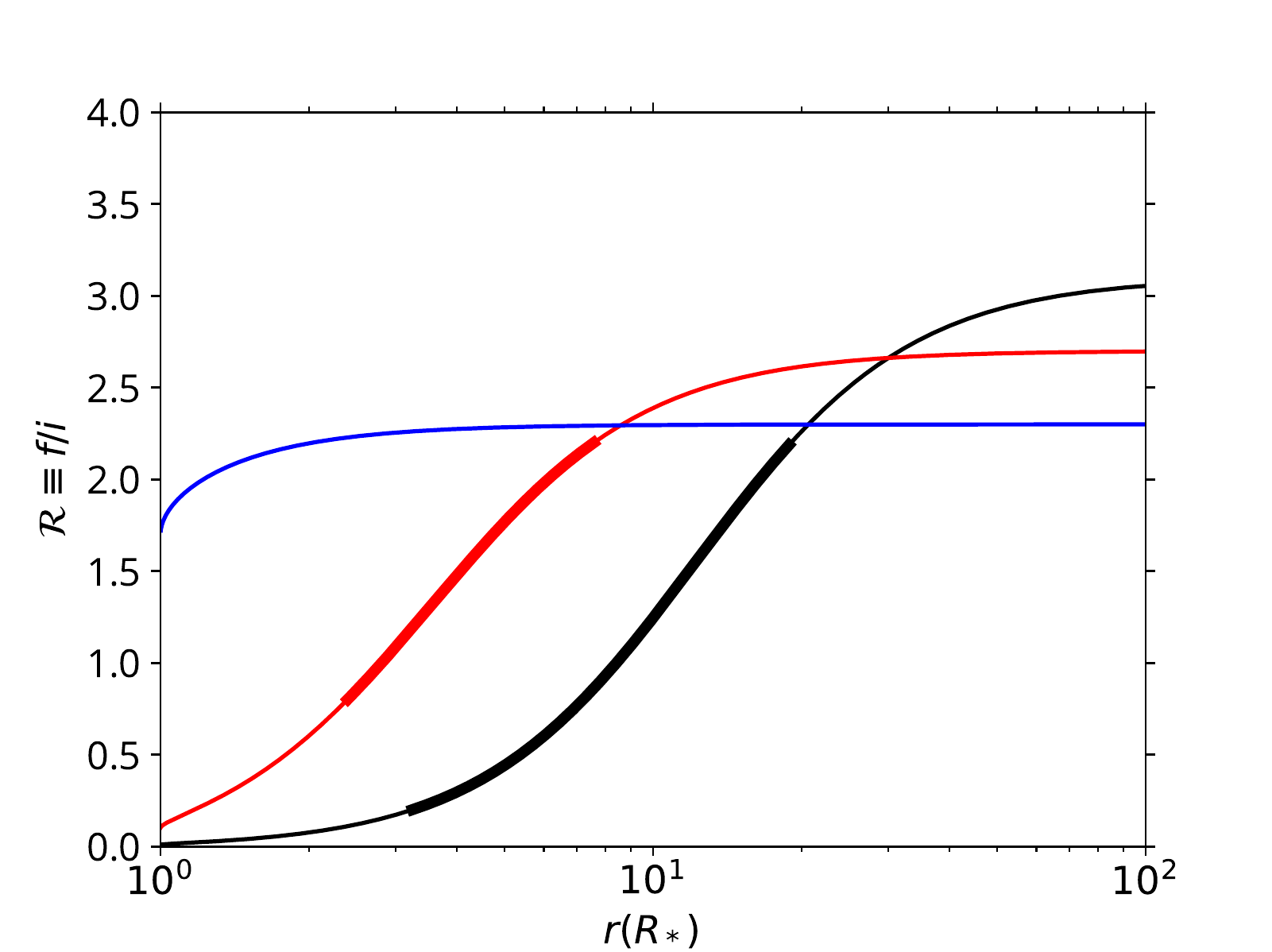}}
\caption{Predicted $R = f/i$ ratio as a function of radius (expressed
  in multiples of stellar radius) for \ion{Ne}{ix} (black),
  \ion{Mg}{xi} (red), and \ion{Si}{xiii} (blue) for a B star with
  similar parameters to HDE~77581. The thickened portions of the
  \ion{Ne}{ix} and \ion{Mg}{xi} curves show the 90\% confidence
  interval for measured values of $R$ and thus for the radius.}
  \label{fig:Rratio}
\end{figure}

In Fig.~\ref{fig:Rratio}, we show predictions for the $R$ ratio as a
function of stellar radius for a B-type star with stellar parameters
similar to HDE~77581 ($T_{\mathrm{eff}} = 25\,000$\,K,
$\log g = 3.0$), using a methodology similar to that described in
Section 2 of \citet{Leutenegger_2006a} to average over the relevant
part of the TLUSTY model stellar continuum, i.e., a non-LTE,
plane-parallel, hydrostatic model stellar atmosphere
\citep{Lanz_2007a}, to estimate the UV field. In the context of the
predicted ratios, and assuming a single radius of formation, the
measurements would imply a formation radius of at least a few stellar
radii thus ruling out a substantial contribution of emission from the
undisturbed stellar wind near the neutron star at
$r \approx 1.7\,R_{\mathrm{HD~77581}}$ for \ion{Ne}{ix} and
\ion{Mg}{xi}.

\subsection{Presence of multi-component medium}

In eclipse ($\phi_{\mathrm{orb}} \approx 0$), co-existing emission
lines from high and low ionization species of several elements have
been previously detected by \citet{Sako_1999a} and
\citet{Schulz_2002b} and in both cases interpreted as signatures of a
multiphase medium, with cool dense clumps co-existing with hotter
plasma.  Similarly, the simultaneous presence of emission lines from
high and low ionized ions of the same element at orbital phase ~0.5,
when the source is strongly absorbed by the accretion wake
(Fig.~\ref{fig:sketch}), implies multiple gas phases with different
temperatures in the medium \citep{Goldstein_2004a,Watanabe_2006a}.

Here, at $\phi_{\mathrm{orb}} \approx 0.25$, we detect K$\alpha$ transitions of L-shell ions of both silicon in emission in the high hardness spectrum, i.e., at higher overall absorption (Sec.~\ref{sec:sihigh}), and of magnesium in absorption in the low hardness spectrum, i.e., at lower overall absorption (Sec.~\ref{sec:Mglow}). In both cases, the L-shell ions co-exist with H- and He-like ions from the same elements. In the case of magnesium, we detect absorption from \ion{Mg}{xi}~r and K$\alpha$ lines from Li-like \ion{Mg}{x}, Be-like \ion{Mg}{ix}, N-like \ion{Mg}{vi}, O-like \ion{Mg}{v} and, tentatively, C-like \ion{Mg}{vii}. These lines imply the presence of cooler material in the system outside of the large scale accretion structure, which should not be crossing our line of sight at $\phi_{\mathrm{orb}} \approx 0.25$ (Fig.~\ref{fig:sketch}), as the comparatively modest measured absorption also implies. Clumps in the structured wind would provide a natural explanation for such material but so would also a structured flow close to the neutron star. While detailed theoretical considerations of the interaction of a clumpy wind with a compact object and its radiation are outside of the scope of this work, we undertake two simplified estimates of whether the observed signatures can be produced in a single-zone absorber assuming a photoionized wind (Sec.~\ref{sec:xstar}) and whether the observed L-shell transitions may come from ions present in the stellar wind itself (Sec.~\ref{sec:powr}).

Further hints of the presence of multiple components come from
the detection of \ion{Ne}{x}~Ly$\beta$ in absorption and
\ion{Ne}{x}~Ly$\alpha$ in emission in the low hardness and thus low
absorption spectrum (Sec.~\ref{sec:Nelow}). Lines of the same element
in both emission and absorption cannot be produced in a single uniform
medium and the observation can only be explained through the presence
of multiple components whose sum results in the observed
spectrum. Both, \ion{Ne}{x}~Ly$\alpha$  and Ly$\beta$ are detected in emission in the high hardness and
thus high absorption spectrum (Sec.~\ref{sec:Nehigh}).

\subsubsection{Photoionization simulations}\label{sec:xstar}

First, to simulate a photoionized wind medium, we approximate the
broadband spectrum of the irradiating continuum as the sum of the
emission from the neutron star and from the companion. Since our data
do not extend above 10\,keV, we rely on the broadband
($\sim$3--70\,keV) fits to \textsl{NuSTAR} observations obtained by
\citet{Fuerst_2014a} for the description of the neutron star
continuum. They analyze two observations and find that a
power-law
model with a Fermi-Dirac cut-off
\citep[\texttt{FDCUT};][]{Tanaka_1986a} gives a good description of
the data:
\begin{equation}
F(E) = A E^{-\Gamma} \left(1 + \exp
  \left(\frac{E-E_{\mathrm{cut}}}{E_{\mathrm{fold}}}\right) \right)^{-1}
\end{equation}
where $A$ is the overall normalization, $\Gamma$ the photon index,
$E_{\mathrm{cut}}$ the cut-off energy , and $E_{\mathrm{fold}}$ the
folding energy. We use $\Gamma = 0.99$, consistent with our
measurement in Sec.~\ref{sec:cont} and with the average $\Gamma$
obtained by \citet{Fuerst_2014a} for their two observations and set
the normalization to 0.277\,keV$^{-1}$\,cm$^{-2}$\,s$^{-1}$ (at
1\,keV), the value we measure. Using their average values, we also
set $E_{\mathrm{cut}} =20.8$\,keV and $E_{\mathrm{fold}}
=11.9$\,keV. We model the contribution of HD~77581, which dominates
the UV-continuum, as a blackbody using the parameters from
\citet{Nagase_1986a}: distance of 2\,kpc, stellar radius of
31\,$R_{\odot}$ and temperature of 25000\,K.

For our purposes here, it is sufficient to determine whether we can
produce the observed range of K$\alpha$ absorption lines of magnesium
ions, from Mg\,{\sc v} to Mg\,{\sc xi}, in a single-zone
absorber. With the above continuum model, we use XSTAR
v2.31\footnote{\url{https://heasarc.gsfc.nasa.gov/lheasoft/xstar/xstar.html}}
to calculate the ionization balance in the illuminated material for
three different densities: $10^{4},$ $10^{10},$ and
$10^{16}$\,cm$^{-3}.$ Using line optical depths from XSTAR, we then
generate approximate synthetic spectra from 8--10\,{\AA}, at the
ionization parameters\footnote{Ionization parameter defined after
  \citet{Tarter_1969a} as $\xi = L/(nr^2)$ with $L$ source luminosity
  above the hydrogen Lyman edge, $n$ the absorbing particle density
  and $r$ the distance from the ionizing source.}  $\xi_1,$ where
\ion{Mg}{v} and \ion{Mg}{xi} have comparable ion fractions, and
$\xi_2$, where \ion{Mg}{v} peaks. Regardless of density, at $\xi_1$ we
overpredict the strength of the \ion{Mg}{xii} line at 8.42\,{\AA}, and
significantly underpredict the \ion{Mg}{v} line. In general, at both
$\xi_1$ and $\xi_2$, we expect lines from \ion{Mg}{vi} to
\ion{Mg}{viii} that are comparable to or stronger than our
\ion{Mg}{v}/\ion{Mg}{xi} lines of interest.

In short, it is difficult to produce both \ion{Mg}{v} and \ion{Mg}{xi}
in a single absorption zone without producing other strong Mg lines
between 8 and 10\,{\AA}. While such lines are not negligible in our
data, they are generally weaker than we would expect based on our
XSTAR calculations; further XSTAR runs show that the discrepancy
between predictions and observations is exacerbated if we suppose that
the absorber is shielded from the stellar continuum and only
illuminated by the neutron star. Thus, although we have not performed
any detailed ionization modeling of the absorber, it is unlikely that
the observed lines originate in a single ionization zone. For the
continuum spectra and ionization parameters considered here, the gas
appears to be thermally stable \citep{Krolik_1981a}, indicating that
multiple distinct absorbers (rather than a multi-phase gas) may be
required.

\subsubsection{Stellar atmosphere modelling}\label{sec:powr}

In a second approach, we use a detailed wind model for the donor star
to see which ions are to be expected and how this changes due to the
presence of a strong X-ray source. We employ the hydrodynamically
consistent stellar atmosphere model calculated for Vela X-1
\citep{Sander_2017b} using the most recent version of the Potsdam
Wolf-Rayet (PoWR) model atmosphere code that is able to obtain the
mass-loss rate and wind stratification self-consistently for a given
set of stellar parameters \citep{Sander_2017a}. This model takes into
account density inhomogeneities in the wind in the form of optically
thin clumps but does not include changes in the wind geometry due to
the presence of the neutron star.

\begin{figure*}
\includegraphics[width=0.49\textwidth]{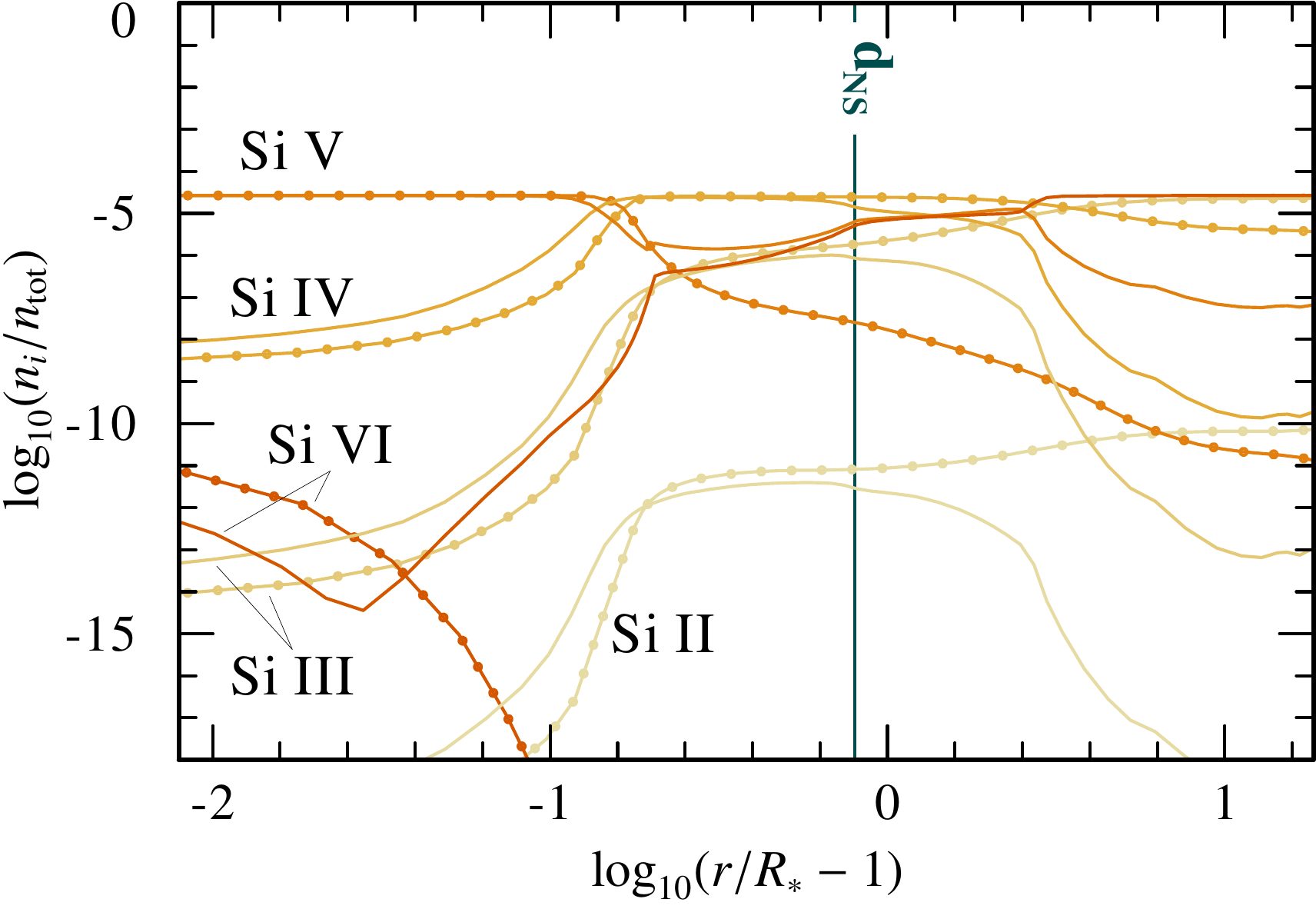}\hfill
\includegraphics[width=0.49\textwidth]{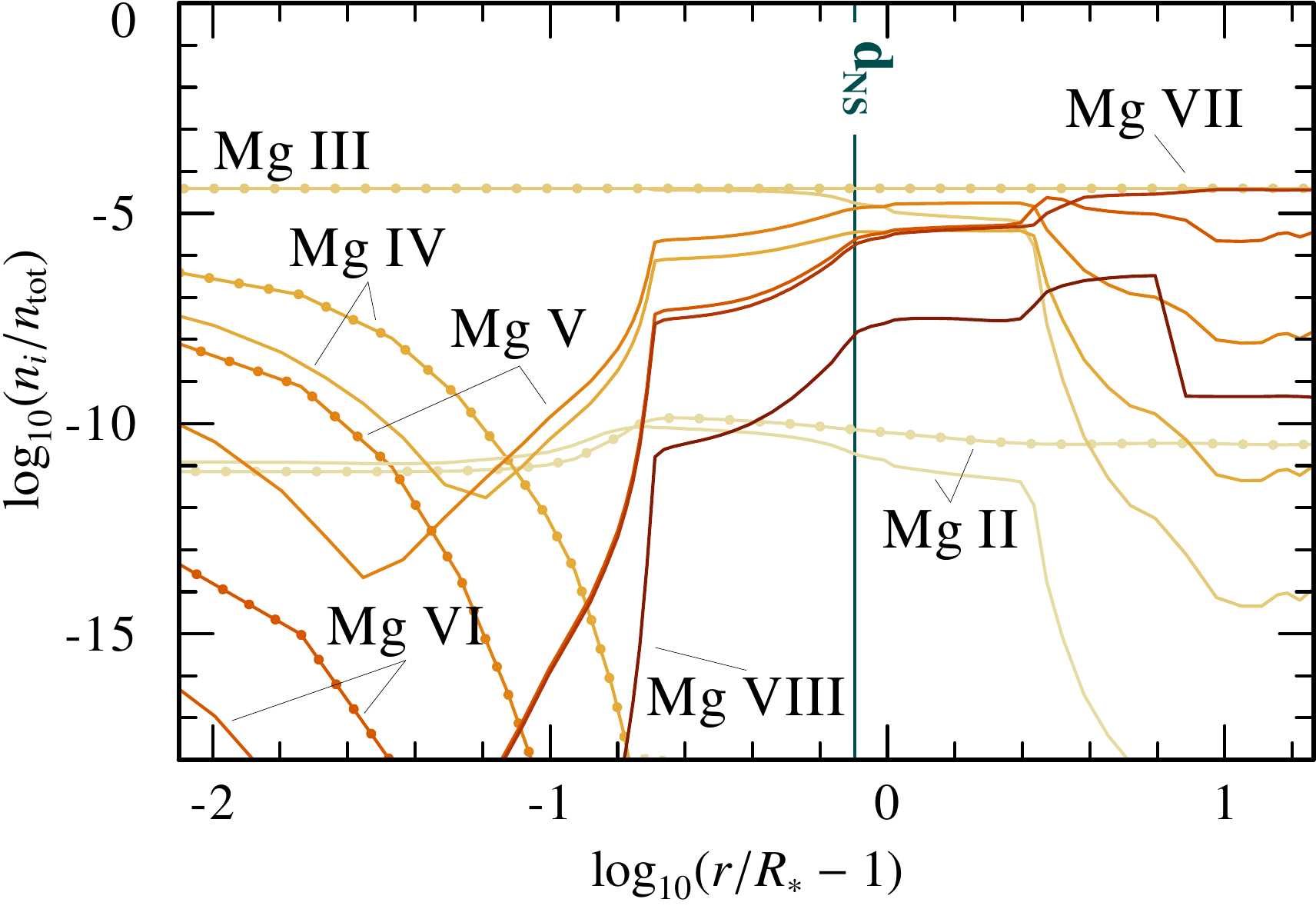}
\caption{Relative population numbers for the ground states of silicon
  (left) and magnesium (right) vs the distance from the donor star in
  two hydrodynamically consistent PoWR models describing the wind of
  HD\,77581. The dotted lines refer to a model without X-ray
  illumination, while the simple lines denote the ionization
  stratification for a model with an X-ray illumination comparable to
  the situation in Vela X-1; same color represents the same ionization
  stage of an element. The vertical line denoted with
  $d_\textsc{ns}$ marks the location of the neutron
  star.}\label{fig:popnum}
\end{figure*}

For the purpose of our work, the donor star models from
\citet{Sander_2017b} were recalculated to account for higher Mg ions
up to \ion{Mg}{viii} and we refer the reader to this work for the
details of the calculations. In these models, Si is covered up to
\ion{Si}{vi}. While we do see higher ions in our X-ray observations,
they would not be expected from a single massive star wind
(Fig.~\ref{fig:popnum}, blue curves).  For the model without X-rays,
only lower ions of both elements are found. For silicon, \ion{Si}{iv}
is leading in the wind until \ion{Si}{iii} takes over in the outer
part. Close to the donor star, \ion{Si}{v} also plays a role, but
already \ion{Si}{vi} is essentially not populated in the wind.  For
magnesium, the situation is even simpler with \ion{Mg}{iii} being the
leading ion in the wind and all higher ions quickly getting
depopulated beyond the surface of the donor.

The situation changes when X-rays are included. Although the treatment
of the X-ray irraditaion in the atmosphere models is just a rough
approximation as these are 1D models and the X-rays are modeled as
Bremsstrahlung, the changes in the population numbers illustrate that
the X-ray illumination due to the accretion onto the neutron star
leads to a more complex structure and to the presence of several
higher ions. \ion{Si}{vi} and \ion{Si}{v} are now only about an order
of magnitude less populated than the leading \ion{Si}{iv}; they remain
important until the wind is essentially transparent and \ion{Si}{vi}
becomes the leading ion in the outer wind (Fig.~\ref{fig:popnum}, left
panel). For magnesium, the situation is roughly similar, but more
complex as more ions are present (Fig.~\ref{fig:popnum}, right
panel). At the distance of the neutron star, the previously
unimportant ions \ion{Mg}{iv} to \ion{Mg}{vii} are now almost
similarly populated. At larger distances, \ion{Mg}{vii} becomes the
leading ion and only \ion{Mg}{vi} remains significantly
populated. Interestingly, the relative population of \ion{Mg}{viii}
increases outwards, but then drops down again. However, the later
could be an artifact from our restriction to 1D models with monotonic
wind velocity laws.

In spite of the limitations of our models, we can conclude that the
ionization situation around Vela~X-1 is complex. The PoWR models have
demonstrated that the donor wind as such produces very low ionized
metals (e.g. \ion{C}{iii}, \ion{Mg}{iii}, \ion{Si}{iv}, \ion{Fe}{iv}),
but the illumination due to X-rays changes this situation
significantly and even in the 1D approach, distinct regions with
different leading ions are present. However, the model also
demonstrated that some ionization stages can exist in the same region.
Given the fact that due to the orbit of the neutron star only a part
of the wind is strongly illuminated, while other parts are less or
even not at all affected by X-rays, it is safe to conclude the
observed X-ray spectrum stems from a multitude of layers with
different ionization situations.

\subsection{Iron fluorescence}\label{sec:ironfluor}

The location of the detected Fe~K$\alpha$ emission features
(Table~\ref{tab:Fe}) is consistent with the K$\alpha_1$ and
K$\alpha_2$ lines of neutral or lowly ionized iron ions below
Fe~\textsc{xii} \citep{House_1969a}. Interestingly, PoWR model
atmosphere calculations for Vela X-1 \citep[][see also Sec.~\ref{sec:powr}]{Sander_2017b} imply that
the main iron ion species present in the wind without X-ray
irradiation should be \ion{Fe}{iv}, followed by \ion{Fe}{iii}. In the
case of strong X-ray irradiation, \ion{Fe}{v} becomes more strongly
populated and surpasses \ion{Fe}{iii}. Furthermore, in this case, also
higher ionization stages (\ion{Fe}{vii} and above) can be reached in
the outer wind.

The flux ratio between Fe~K$\beta$ and Fe~K$\alpha$
is consistent with the theoretical expectations of $\sim$0.13--0.14
\citep{Palmeri_2003a}. We note, however, the large uncertainties on
Fe~K$\beta$ which are exacerbated by the presence of the Fe~K edge.

\citet{Kallman_2004a} predict the curve of growth for the equivalent
width of Fe~K$\alpha$, $\mathrm{EQW}(\mathrm{Fe~K}\alpha)$, in units
of eV, and the hydrogen equivalent absorption column density of the
reprocessing material, $N^{22}_{\mathrm{H}}$, in units of
10$^{22}$\,cm$^{-2}$, to be
$\mathrm{EQW}(\mathrm{Fe~K}\alpha) \backsimeq 3 N^{22}_{\mathrm{H}}$
\citep[see also][]{Inoue_1985a}.  It has been observationally verified
for HMXBs by \citet{Torrejon_2010a} using a \textsl{Chandra}-HETG
sample that included Vela X-1. To compare our results, we calculate
the equivalent width of the Fe~K$\alpha$ and K$\beta$ lines in eV
(Table~\ref{tab:Fe}) and use the values for $N_{\mathrm{H}}$ obtained
from continuum fits (Sec.~\ref{sec:cont}). The observed lines are much
stronger than would be predicted by the above correlation; however,
\citet{Torrejon_2010a} obtain similar results for the same
  data: in their analysis, Vela X-1 clearly lies above the
correlation.

Several effects can contribute to the disagreement between the
measured values and the theoretically predicted curve of growth: on
one hand, $N_{\mathrm{H}}$ measurements are generally dependent on the
continuum model and our choice of a power law to describe the
2.5--10\,keV may lead to systematic deviations, although we note that
our results roughly agree with the time averaged analyses of the same
\chandra observation by other authors \citep{Goldstein_2004a,
  Torrejon_2010a}. On the other hand, \citet{Kallman_2004a} compute
the above relationship for a spherical geometry that is clearly not
given in Vela~X-1, given the influence of the compact object on the
large-scale wind structure. The surface of the companion star can also
contribute significantly to the observed FeK$\alpha$ emission as shown
by \citet{Watanabe_2006a} who, however, also show that this cannot
account for the full observed equivalent width. We further implicitly
assume that the iron fluorescence happens in the same material that is
absorbing the X-rays along the line of sight. For a clumpy absorber
this does not have to be the case and the neutron star may be hidden
behind some material without the ambient source of fluorescent
emission being affected \citep{Inoue_1985a}.  A cool cloud, for
example the accretion wake, irradiated by the neutron star could
account for the excess emission \citep{Watanabe_2006a}. Three
reprocessing sites --- stellar wind, companion atmosphere and cold
material in the proximity of the neutron star --- have been also
suggested by \citet{Sato_1986a}.
We note that such a geometry would imply a more complex continuum
model, with multiple differently absorbed continua as has been
employed, e.g., by \citet{Martinez-Nunez_2014a} to describe
\emph{XMM-Newton}-pn observations of Vela~X-1, but is not useable in
our case due to data quality.

\section{Summary and outlook}\label{sec:sum}

We have shown that at $\phi_{\mathrm{orb}} \approx 0.25$, when our
line of sight to the compact object is expected to be least affected
by the large-scale accretion structure, Vela X-1 shows spectral
changes on timescales of $\sim$hours ($\sim$a few ks). These changes
are at least partially driven by fluctuating absorption. The variable
absorption cannot be explained by unperturbed clumpy stellar winds,
but the large-scale accretion structure (accretion wake,
photoionization wake, accretion stream) is also not expected to vary
on such short timescales or to cross our line of sight at this orbital
phase at all. 

We have detected variable emission and absorption line features from
neon, magnesium and silicon at $\phi_{\mathrm{orb}} \approx 0.25$,
i.e., at an orbital phase when few line features have been reported
previously. Overall, the observational picture is consistent with a
complex geometry, with multiple phases co-existing in the system and
contributing different components to the overall observed
spectrum. He-like triplet diagnostics imply that a substantial
fraction of the material along the line of sight further away from the
compact object is ionized to He-like charge state of magnesium and
neon. The L-shell ions of Si and Mg are a signature of the presence of
cooler material. The strong Fe~K$\alpha$ implies the presence of cool
material as source for the fluorescent emission.

We speculate that dense regions could arise in shocks close to the
compact object and have longer crossing times and higher equivalent
hydrogen column density than unperturbed wind clumps, thus
explaining the observed variability. The possible origin of such dense regions through the interaction of wind clumps at shocks is a notion to be explored in upcoming simulations of wind accretion in HMXBs that can then be compared to the observations
presented here.

We have shown that time-resolved spectroscopy is crucial to
disentangle the contributions of different components. However, the
presented \textsl{Chandra}-HETG observation does not allow us,
for example, to explore possible fluctuations in absorption during the
high hardness periods, to perform pulse-resolved spectroscopy of the
line features or to measure the possible response of the ionized gas
to changes in the X-ray flux of the neutron star. More observations at
$\phi_{\mathrm{orb}} \approx 0.25$ will allow to accumulate more data
and give insight into the finer structure of the structured absorber
but the major leap forward will only be possible with future
intruments that have a higher effective area. Future work would also profit from an improvement in line reference values of L-shell ions for magnesium, both to enable a better analysis of the currently available high resolution data and for forthcoming mission with their higher throughput and/or higher resolution.

In particular, Vela X-1 was among the planned targets for
\textsl{Astro-H}/\textsl{Hitomi} \citep{Kitamoto_2014a} and is thus
also likely to be among the main sources observed in the future by
approved microcalorimeter missions such as \emph{Athena}
\citep{Nandra_2013a} and the X-ray Astronomy Recovery Mission
(\emph{XARM})\footnote{\url{https://heasarc.gsfc.nasa.gov/docs/xarm/}}
as well as planned high-resolution spectroscopy grating missions such
as \textsl{Arcus} \citep{Smith_2014a}, a Medium-Class Explorer mission
that has recently been selected for a concept study, and X-ray
Surveyor/\emph{LYNX}\footnote{\url{https://wwwastro.msfc.nasa.gov/lynx/}}.
Their higher sensitivity will allow a detailed study of changing
absorption -- but our results also imply that we will need to
carefully consider ks-timescale variability when drawing any
conclusions from these future observations.

\begin{acknowledgements} VG is thankful for support of her work
  through the ESA internal research fellowship. Part of this work was
  performed under the auspices of the U.S. Department of Energy by
  Lawrence Livermore National Laboratory under Contract
  DE-AC52-07NA27344. Support for this work was provided by NASA
  through the Smithsonian Astrophysical Observatory (SAO) contract
  SV3-73016 to MIT for Support of the Chandra X-Ray Center (CXC) and
  Science Instruments; CXC is operated by SAO for and on behalf of
  NASA under contract NAS8-03060. IEM acknowledges funding from the
  European Union’s Horizon 2020 research and innovation programme
  under the Marie Skłodowska-Curie grant agreement No 665501 with the
  Research Foundation Flanders (FWO). AACS is supported by the
  Deutsche Forschungsgemeinschaft (DFG) under grant HA 1455/26.  JN
  acknowledges support from NASA through the Hubble Postdoctoral
  Fellowship Program, grant HST-HF2-51343.001-A. SMN acknowledges
  support by research project ESP2016-76683-C3-1-R. For the initial
  data exploration, this research has made use of the \chandra
  Transmission Grating Data Catalog and
  Archive\footnote{\url{http://tgcat.mit.edu/}}
  \citep[\texttt{tgcat};][]{Huenemoerder_2011a}. This research has
  made use of ISIS functions (\texttt{isisscripts})\footnote{\url{http://www.sternwarte.uni-erlangen.de/isis/}}
  provided by ECAP/Remeis observatory and MIT and of NASA's
  Astrophysics Data System Bibliographic Service (ADS).  We thank John
  E.  Davis for the development of the
  \texttt{slxfig}\footnote{\url{http://www.jedsoft.org/fun/slxfig/}}
  module used to prepare most of the figures in this work. Some of the
  color schemes used were based on Paul Tol's palettes and
  templates\footnote{\url{https://personal.sron.nl/~pault/}}. IEM,
  AACS, FF, PK, MK, SMN and JW are grateful for the hospitality of the
  International Space Science Institute (ISSI), Bern, Switzerland
  which sponsored a team meeting initiating a tighter collaboration
  between stellar wind and X-ray specialists.  And finally, we thank
  the anonymous referee whose comments and suggestions helped us to
  improve this manuscript.
\end{acknowledgements}


\end{document}